\documentclass[11pt]{article}

\usepackage[final]{acl}

\usepackage{times}
\usepackage{latexsym}
\usepackage{amsmath}
\usepackage{booktabs}
\usepackage{multirow}
\usepackage{booktabs}
\usepackage{multirow}
\usepackage{booktabs}
\usepackage{tabularx}
\usepackage{array}
\usepackage{multirow}
\usepackage{graphicx}
\usepackage{adjustbox}
\usepackage{subcaption}
\usepackage{makecell}
\usepackage{multirow}
\usepackage{makecell}
\usepackage[table]{xcolor}
\usepackage{booktabs}

\newcolumntype{Y}{>{\centering\arraybackslash}X}
\newcolumntype{C}[1]{>{\centering\arraybackslash}p{#1}}

\usepackage{amsmath}
\usepackage{etoolbox}

\usepackage{amssymb}

\usepackage[T1]{fontenc}

\usepackage[utf8]{inputenc}

\usepackage{microtype}

\usepackage{inconsolata}

\usepackage{graphicx}

\title{State-Anchored Complete-View Distillation for Robust Conversational Multimodal Emotion Recognition}

\author{
\textbf{Zhaoyan Pan}\textsuperscript{1,*},
\textbf{Xiangdong Li}\textsuperscript{1,*},
\textbf{Wenke Wu}\textsuperscript{1,*},
\textbf{Mengting Ma}\textsuperscript{2}, \\
\textbf{Ye Lou}\textsuperscript{1},
\textbf{Ji Zhou}\textsuperscript{1},
\textbf{Jiatong Pan}\textsuperscript{1},
\textbf{Wei Zhang}\textsuperscript{1,2,$\dagger$} \\
\textsuperscript{1}Shcool of Software Technology, Zhejiang University \\
\textsuperscript{2}Shcool of Computer Science and Technology, Zhejiang University \\
\textsuperscript{1}\texttt{\{zhaoyanpan, xiangdong.li, wenkewu, ye\_lou, jizhou, jiatongpan\}@zju.edu.cn}\\
\textsuperscript{2}\texttt{\{mtma, ctszhangwei\}@zju.edu.cn}\\
\textsuperscript{*}Equal contribution.
\textsuperscript{$\dagger$}Corresponding author.
}

\begin{document}
\maketitle 
\begin{abstract}
Conversational multimodal emotion recognition (MER) requires reliable prediction when language, acoustic, or visual observations are missing or unreliable.
Many missing-modality methods reconstruct absent inputs, yet such recovery can be non-unique in dialogue context, and nonverbal cues may conflict with the target utterance.
To this end, we propose \textbf{CoRe-KD} (\textbf{Co}mplete-view \textbf{Re}ference-guided \textbf{K}nowledge \textbf{D}istillation), a state-anchored, conflict-regularized complete-view distillation framework for robust conversational MER.
A complete-view teacher provides structured references, including prediction-level references, fused states, and modality-specific states.
\textbf{Complete-view State Anchoring} (CSA) aligns incomplete-view student predictions and states with these references, while \textbf{Nonverbal Conflict Exposure} (NCE) trains on target-preserving nonverbal conflict views to reduce donor-label bias.
Experiments on IEMOCAP and MELD, with CMU-MOSEI as a supplementary utterance-level check, show consistent gains under fixed- and random-missing protocols.
Comprehensive ablation studies and further analyses support the role of CSA and the complementary effect of NCE.

\end{abstract}

\section{Introduction}

Conversational multimodal emotion recognition (MER) infers utterance-level emotions from dialogue context and language, acoustic, and visual observations~\citep{busso2008iemocap, poria2019meld}.
In practical scenarios, MER models often face incomplete observations, where any modality may be unavailable or partially observed, yielding insufficient evidence under fixed-missing or random-missing modality settings~\citep{zhao2021mmin,lian2023gcnet}.
Conversational nonverbal signals can also be loosely aligned with the target utterance due to speaker ambiguity, delayed reactions, or temporal mismatch~\citep{tsai2019mult}.

\begin{figure}[t]
    \centering
    \includegraphics[width=\linewidth]{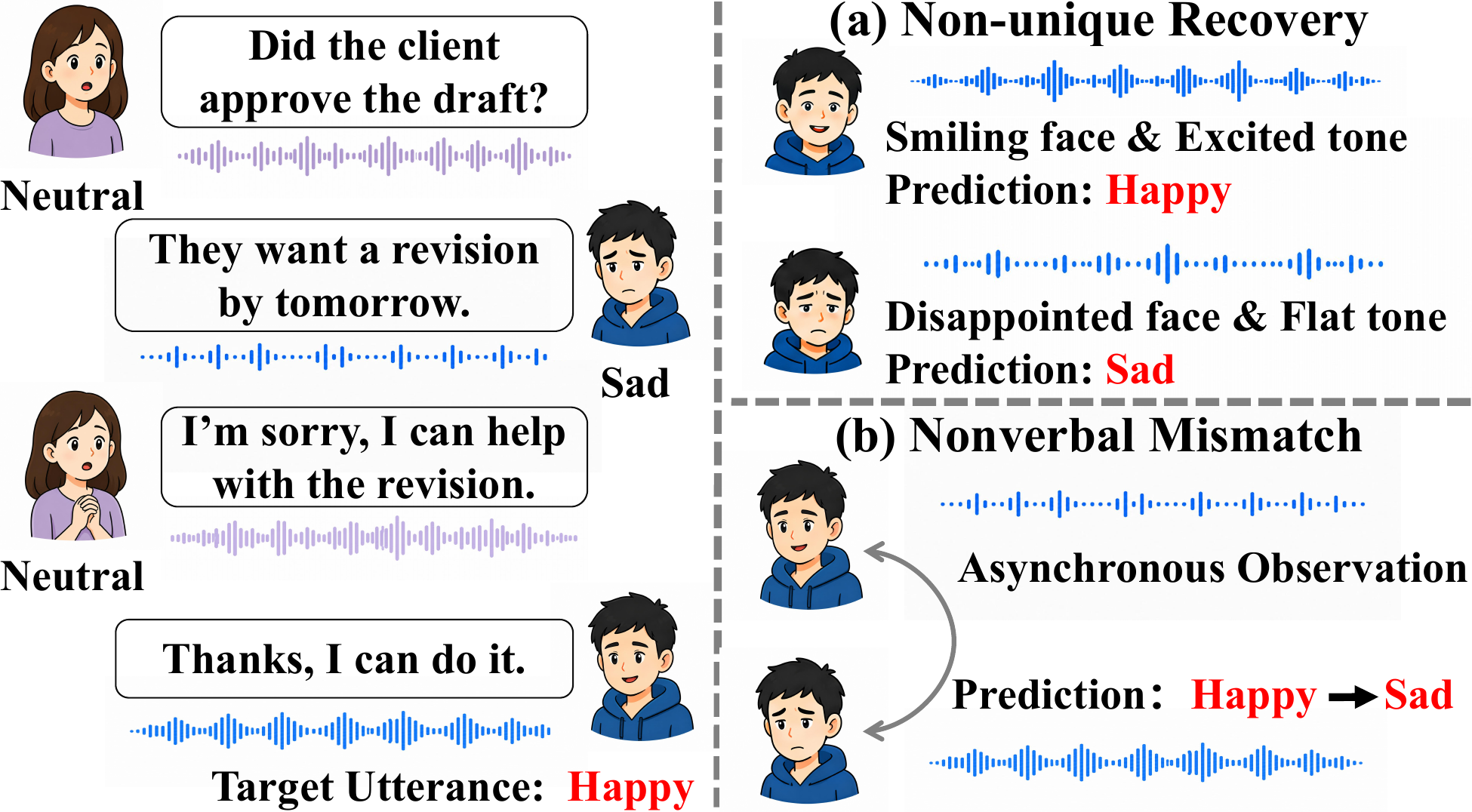}
    \caption{
    Panel (a) shows two issues: missing nonverbal cues lead to \emph{non-unique recovery}, as the same utterance may admit multiple audio-visual completions; recovery or output-level distillation may leave \emph{complete-view evidence underspecified} at fused and modality-specific levels.
    Panel (b) shows \emph{target-inconsistent nonverbal evidence}, where asynchronous or mismatched cues may contradict the target utterance.
    }
    \label{fig:intropic}
\end{figure}

Recent robust MER methods have made progress in incomplete-input learning through reconstruction, distillation, and reliability-aware learning.
Reconstruction-based methods estimate unavailable modality representations from observed inputs~\citep{pham2019found, yuan2021transformer, zhao2021mmin,guo2024mplmm, zhang2024lnln}.
Distillation- and alignment-based methods transfer complete-view predictions, representations, correlations, relations, or category-level knowledge to incomplete-view models~\citep{wang2020kd, lin2023missmodal, li2023dmd, li2024corrkd, ma2024sdt, zhuang2025cmad}.
Reliability-, modality-, and expert-aware methods estimate confidence, uncertainty, quality, or modality-specific knowledge under partial modalities~\citep{han2022tmc, xu2024momke, bi2025dream, dai2025umm, he2026comp, miyoshi2026dqf, zhuang2026mcur}.
These methods provide useful supervision for reconstructing unavailable information, transferring complete-view knowledge, or deciding how to weight and route observed modalities.

However, as illustrated in Figure~\ref{fig:intropic}, incomplete and unreliable conversational observations still expose three issues.
\emph{(i) Non-unique recovery}: Panel~(a) shows that when nonverbal cues are missing, the same language context can be compatible with multiple plausible audio-visual completions, making raw or feature-level recovery ambiguous.
\emph{(ii) Underspecified complete-view evidence}: reconstruction or output-level distillation may provide useful guidance, but may not explicitly specify which complete-view fused evidence and modality-specific evidence an incomplete-view student should preserve.
\emph{(iii) Target-inconsistent nonverbal evidence}: Panel~(b) shows that observed audio and visual cues may be temporally asynchronous or inherently mismatched with the target utterance, introducing unreliable nonverbal evidence.
These issues motivate complete-view references that anchor incomplete-view states at both fused and modality-specific levels instead of enforcing a unique raw completion, together with target-preserving conflict exposure for unreliable nonverbal observations.

To address these issues, we propose \textbf{CoRe-KD}
(\textbf{Co}mplete-view \textbf{Re}ference-guided \textbf{K}nowledge \textbf{D}istillation),
a state-anchored and conflict-regularized complete-view distillation framework for robust conversational MER.
Different from methods that reconstruct missing inputs or mainly transfer teacher outputs, CoRe-KD treats the complete view as a structured task reference: predictive distributions guide incomplete-view decisions, fused Gaussian-inspired anchors preserve complete-view evidence, and modality-specific states serve as references for unavailable-modality-state alignment.
Inspired by Product-of-Experts (PoE) modeling and multimodal PoE inference~\citep{hinton2002training,wu2018multimodal}, these location-scale states provide precision-weighted fusion across different modality subsets.
CoRe-KD then uses \textbf{Complete-view State Anchoring} (CSA) to align incomplete-view predictions, fused states, and unavailable-modality states with the complete-view teacher, avoiding the need to identify a unique raw completion (issue i) while explicitly preserving fused and modality-specific evidence (issue ii); it further uses \textbf{Nonverbal Conflict Exposure} (NCE) to train on target-preserving nonverbal conflict views, mitigating sensitivity to target-inconsistent nonverbal cues under controlled conflict views (issue iii).   
The main contributions of our paper can be summarized as follows:
\begin{itemize}
    \item We propose \textbf{CoRe-KD}, a novel complete-view distillation framework that avoids input reconstruction by leveraging a complete-view teacher to provide three-level supervisory signals for robust conversational MER.



    \item We introduce two key components of CoRe-KD: \textbf{Complete-view State Anchoring} (CSA), which aligns incomplete-view predictions and teacher-provided anchors at both fused and modality-specific levels, and \textbf{Nonverbal Conflict Exposure} (NCE), which uses target-preserving nonverbal conflict views as a regularization signal to reduce the sensitivity to unreliable nonverbal observations.
    
    \item Extensive experiments on IEMOCAP and MELD demonstrate consistent improvements under fixed- and random-missing protocols, while CMU-MOSEI provides a supplementary utterance-level check.
    
\end{itemize}

\section{Related Work}

\paragraph{Multimodal Affective Computing.}
Multimodal affective computing models sentiment, emotion, and social-affective meaning from language, audio, vision, and context~\citep{busso2008iemocap,poria2019meld}.
Classical studies explore multimodal fusion and cross-modal interaction through tensor fusion, low-rank fusion, memory fusion, Transformer-based alignment, and pretrained language model integration~\citep{zadeh2017tensor,liu2018efficient,zadeh2018memory,tsai2019mult,rahman2020integrating}.
For conversational scenarios, context-aware emotion recognition models dialogue history and speaker interactions to disambiguate utterance-level affect~\citep{hazarika2018icon,majumder2019dialoguernn}.
Recent studies improve affective representation through correlation modeling, semantic constraints, and context-dependent interaction~\citep{xu2025towards,pan2026isolatedutterancescueguidedinteraction}, while robust multimodal learning studies prediction under unavailable, degraded, or unreliable observations~\citep{liang2021multibench,pham2019found,zhao2021mmin}.
Related sarcasm and misinformation studies highlight semantic consistency, evidence reasoning, and robustness to noisy or incomplete cues~\citep{wei2024g,wei2024towards,yuan2025enhancing,wei2025deepmsd,zhou2025ldgnet,zhou2025towards,zhou2026diverdynamiciterativevisual}.
Together, these motivate target-consistent affective evidence recovery under missing, misaligned, or unreliable nonverbal observations in conversational MER.

\paragraph{Robust Multimodal Emotion Recognition.}
Robust multimodal learning studies prediction under unavailable, degraded, or unreliable observations~\citep{liang2021multibench}, while conversational MER focuses on utterance-level emotion recognition in dialogue, with IEMOCAP~\citep{busso2008iemocap} and MELD~\citep{poria2019meld} as common benchmarks.
Apart from complete-view distillation discussed below, existing methods mainly follow two lines.
Reconstruction-based methods recover missing information, from early cross-modal translation or imagination methods in multimodal affective learning such as MCTN~\citep{pham2019found} and MMIN~\citep{zhao2021mmin} to recent restoration-, diffusion-, prompting-, and proxy-based methods such as IMDer~\citep{wang2023imder}, MPLMM~\citep{guo2024mplmm}, and LNLN~\citep{zhang2024lnln}.
Recent reliability- and modality-aware robust multimodal methods further improve incomplete-view learning with modality experts, dynamic modality enhancement, cross-modal prompting, or uncertainty-regularized fusion, as represented by MoMKE~\citep{xu2024momke}, DREAM~\citep{bi2025dream}, ComP~\citep{he2026comp}, and MCUR~\citep{zhuang2026mcur}.
These studies improve robustness when modalities are unavailable or incomplete.
In dialogue, however, raw recovery can be non-unique, and nonverbal signals may be target-inconsistent.

\begin{figure*}[t]
    \centering
    \includegraphics[width=0.98\textwidth]{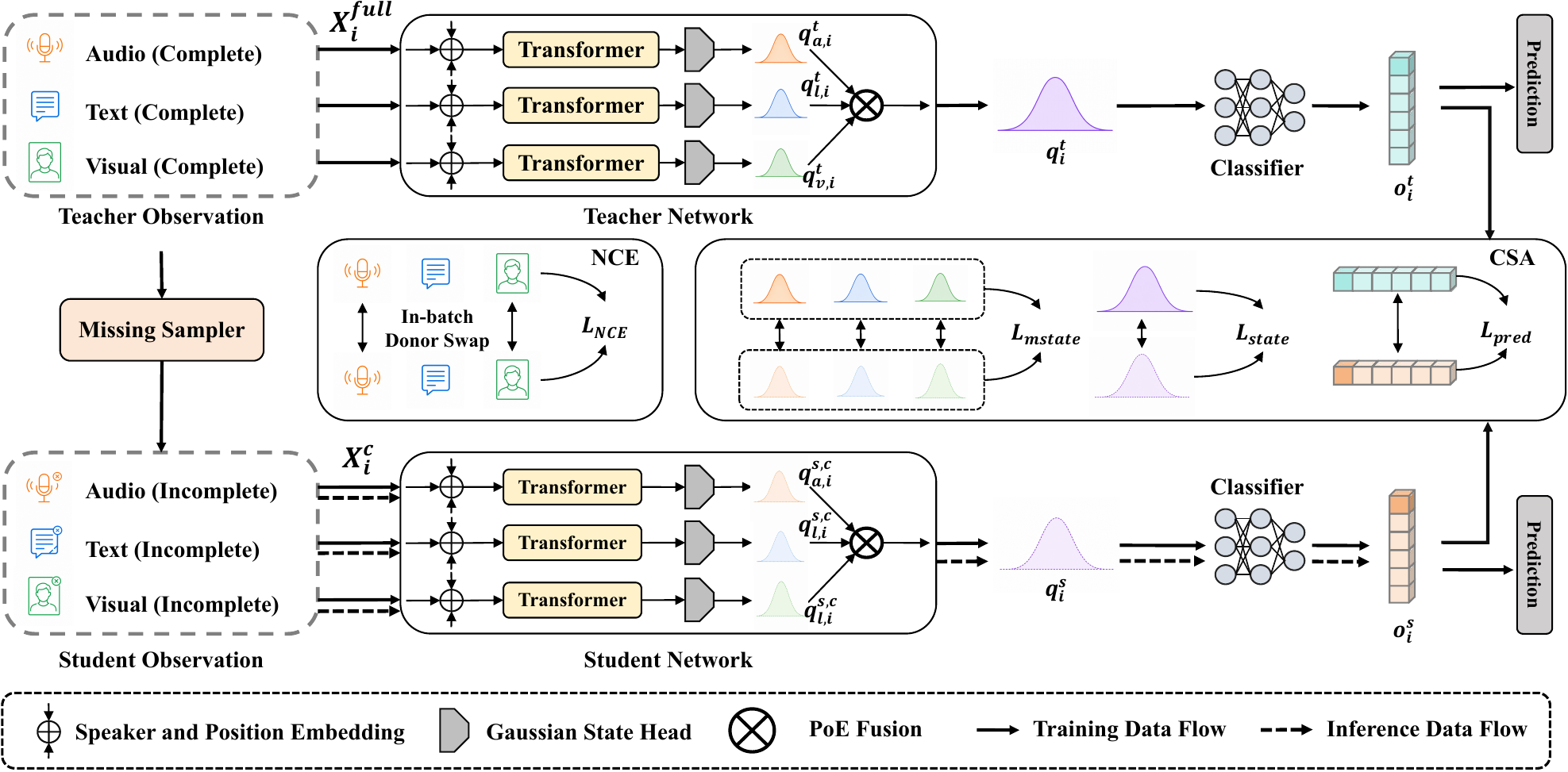}
    \caption{
    \textbf{Overall framework of CoRe-KD.}
    A frozen complete-view teacher provides prediction-, fused-state-, and modality-specific-state references for an incomplete-view student; CSA aligns these references to preserve complete-view evidence, NCE constructs target-preserving audio/visual conflict views via in-batch donor swaps, and only the student is retained at inference.
    }
    \label{fig:framework}
\end{figure*}

\paragraph{Knowledge Distillation for Incomplete Multimodal Affective Computing.}
Knowledge distillation has been widely used in multimodal affective learning to transfer knowledge from complete, stronger, or teacher models to students under incomplete or heterogeneous inputs.
Early incomplete-modality distillation uses complete-view teachers to guide missing-input students~\citep{wang2020kd}.
Recent methods enrich the transferred signals with missing-modality supervision, decoupled factors, self-distillation, correlation knowledge, or modality-aware distillation, as in MissModal~\citep{lin2023missmodal}, DMD~\citep{li2023dmd}, UMDF~\citep{li2024umdf}, Corr-KD~\citep{li2024corrkd}, and CMAD~\citep{zhuang2025cmad}.
In conversational MER, SDT uses self-distillation for multimodal fusion~\citep{ma2024sdt}, while TelME uses teacher-guided learning for modality interaction and dialogue-context modeling~\citep{yun2024telme}.
These methods provide complete-view supervision at the output, representation, relation, correlation, or category level.
Under condition-specific observations, fused and modality-specific complete-view states may remain implicit.

\section{Preliminaries}

\subsection{Problem Formulation}

Given a dialogue \(\mathcal D=(u_1,\ldots,u_N)\), conversational MER predicts the emotion label
\(y_i\in\mathcal Y\) of target utterance \(u_i\) from context \(\mathcal D_i\) and multimodal observations over
\(\mathcal M=\{\mathrm{\ell},\mathrm{a},\mathrm{v}\}\).
An observation condition \(c\in\mathcal C\) specifies a modality-availability pattern.
Under condition \(c\), only modalities
\(\mathcal A_i^c\subseteq\mathcal M\) are accessible:
\begin{equation}
\begin{aligned}
\mathbf X_i^c
&=
\{\mathbf x_i^{m,c}\mid m\in\mathcal A_i^c\},\\
\hat y_i^c
&=
\arg\max_{y\in\mathcal Y}
p_\theta(y\mid \mathbf X_i^c,\mathcal D_i).
\end{aligned}
\end{equation}
The complete view has \(\mathcal A_i^c=\mathcal M\); otherwise the view is incomplete.
Target-preserving nonverbal conflict views are defined in Section~\ref{sec:NCE}.

\subsection{Gaussian-inspired State Fusion}
\label{sec:state_fusion}

\paragraph{Gaussian-inspired State.}
We use a Gaussian-inspired state as a compact location-scale representation for modality fusion and teacher-reference matching~\citep{wu2018multimodal}.
It is used as a practical matching state rather than a calibrated probabilistic posterior.
Let
\(\mathcal G(\boldsymbol\mu,\boldsymbol\sigma)=
\mathcal N(\boldsymbol\mu,\operatorname{diag}(\boldsymbol\sigma^2))\).
For a generic modality representation \(\mathbf h_{m,i}^{c}\), a state head produces a diagonal Gaussian-inspired state and its precision:
\begin{equation}
\begin{aligned}
\left(
q_{m,i}^{c},
\boldsymbol\kappa_{m,i}^{c}
\right)
&=
\operatorname{StateHead}_{m}
\left(
\mathbf h_{m,i}^{c}
\right),\\
q_{m,i}^{c}
&=
\mathcal G
\left(
\boldsymbol\mu_{m,i}^{c},
\boldsymbol\sigma_{m,i}^{c}
\right).
\end{aligned}
\end{equation}
Here, \(\boldsymbol\kappa_{m,i}^{c}=(\boldsymbol\sigma_{m,i}^{c})^{-2}\) is the diagonal precision used for fusion.
The implementation of \(\operatorname{StateHead}_{m}\), including log-variance parameterization and clipping, is provided in Appendix~\ref{app:teacher_model}.

\paragraph{Product-of-Experts Fusion.}
Given the accessible modality set \(\mathcal A_i^c\), available modality states are fused by a Product-of-Experts rule:
\begin{equation}
\label{eq:poe_fusion}
\begin{aligned}
(\boldsymbol\sigma_i^c)^2
&=
\left(
\mathbf 1+
\sum_{m\in\mathcal A_i^c}
\boldsymbol\kappa_{m,i}^{c}
\right)^{-1},\\
\boldsymbol\mu_i^c
&=
(\boldsymbol\sigma_i^c)^2
\odot
\sum_{m\in\mathcal A_i^c}
\boldsymbol\kappa_{m,i}^{c}
\odot
\boldsymbol\mu_{m,i}^{c}.
\end{aligned}
\end{equation}
The \(\mathbf 1\) term corresponds to a zero-mean unit-precision prior and stabilizes fusion when few modalities are available.
The fused state is denoted as
\(q_i^c=\mathcal G(\boldsymbol\mu_i^c,\boldsymbol\sigma_i^c)\).

\section{Method}
\label{sec:method}

\subsection{Overview}

\textbf{CoRe-KD} follows a complete-to-incomplete training pipeline for robust conversational MER.
A frozen complete-view teacher first encodes clean language, acoustic, and visual observations to produce prediction- and state-level references, while the student processes condition-specific views sampled from the same dialogue inputs.
CSA anchors the student's predictions and states to the teacher reference, and NCE adds target-preserving nonverbal conflict views during training; only the student is retained for inference.

\subsection{Reference State Estimation}

For compact notation, we omit the dialogue context \(\mathcal D_i\) in encoder equations.
The generic representation \(\mathbf h\) in Section~\ref{sec:state_fusion} is instantiated by student encoder outputs: for each accessible modality \(m\in\mathcal A_i^c\), the student encodes \(\mathbf x_i^{m,c}\) into the target-utterance representation \(\mathbf h_{m,i}^{s,c}\) and maps it to a state \(q_{m,i}^{s,c}\) with precision \(\boldsymbol\kappa_{m,i}^{s,c}\):
\begin{equation}
\begin{aligned}
\mathbf h_{m,i}^{s,c}
&=
E_m^s
\left(
\mathbf x_i^{m,c}
\right),\\
\left(
q_{m,i}^{s,c},
\boldsymbol\kappa_{m,i}^{s,c}
\right)
&=
\operatorname{StateHead}_{m}^{s}
\left(
\mathbf h_{m,i}^{s,c}
\right).
\end{aligned}
\end{equation}
The accessible student-side states are fused by Eq.~\eqref{eq:poe_fusion}, yielding
\(q_i^{s,c}=\mathcal G(\boldsymbol\mu_i^{s,c},\boldsymbol\sigma_i^{s,c})\).

On the clean complete view, the frozen teacher uses the same state interface and provides
\begin{equation}
\mathcal T_i
=
T(\mathbf X_i^{\mathrm{full}},\mathcal D_i)
=
\left(
\mathbf o_i^t,
p_i^t,
q_i^t,
\{q_{m,i}^{t}\}_{m\in\mathcal M}
\right),
\end{equation}
where \(\mathbf X_i^{\mathrm{full}}\) is the clean complete-view input; \(\mathbf o_i^t\), \(p_i^t=\operatorname{softmax}(\mathbf o_i^t)\), \(q_i^t\), and \(\{q_{m,i}^{t}\}_{m\in\mathcal M}\) denote the teacher logits, predictive distribution, fused state, and modality-specific states, respectively.
The teacher implementation is detailed in Appendix~\ref{app:teacher_model}; teacher and student share the same state parameterization but not parameters.

Student-side logits and predictive distribution are
\(\mathbf o_i^{s,c}=C^s(\boldsymbol\mu_i^{s,c})\) and
\(p_i^{s,c}=\operatorname{softmax}(\mathbf o_i^{s,c})\).
We use \(\mathcal K_\tau\) for prediction matching and \(\mathcal D_G\) for state matching:
\begin{equation}
\begin{aligned}
\pi_\tau(\mathbf o)
&=
\operatorname{softmax}
\left(
\mathbf o/\tau
\right),\\
\mathcal K_\tau(\mathbf o^t,\mathbf o^s)
&=
\tau^2
D_{\mathrm{KL}}
\left(
\pi_\tau(\mathbf o^t)
\|
\pi_\tau(\mathbf o^s)
\right).
\end{aligned}
\end{equation}
For two diagonal states
\(q=\mathcal G(\boldsymbol\mu,\boldsymbol\sigma)\)
and
\(q'=\mathcal G(\boldsymbol\mu',\boldsymbol\sigma')\), we define
\begin{equation}
\mathcal D_G(q,q')
=
\frac{1}{d}
\left(
\|\boldsymbol\mu-\boldsymbol\mu'\|_2^2
+
\|\boldsymbol\sigma-\boldsymbol\sigma'\|_2^2
\right),
\end{equation}
where \(d\) is the state dimension.
This squared \(2\)-Wasserstein form provides the state distance for teacher-reference matching.
\subsection{Complete-view State Anchoring}

CSA anchors three student outputs to the complete-view teacher reference \(\mathcal T_i\): predictions, fused states, and unavailable-modality states.

\paragraph{Prediction anchoring.}
CSA combines supervised classification with complete-view prediction distillation, so incomplete-view predictions remain guided by the complete-view decision:
\begin{equation}
\begin{aligned}
\mathcal L_{\mathrm{pred}}^{i,c}
&=
\operatorname{CrossEntropy}
\left(
y_i,
\mathbf o_i^{s,c}
\right)\\
&\quad+
\lambda_{\mathrm{kd}}
\mathcal K_\tau
\left(
\mathbf o_i^t,
\mathbf o_i^{s,c}
\right).
\end{aligned}
\end{equation}

\paragraph{Fused-state anchoring.}
Beyond prediction-level supervision, CSA aligns the student fused state with the complete-view teacher state:
\begin{equation}
\mathcal L_{\mathrm{state}}^{i,c}
=
\mathcal D_G
\left(
q_i^{s,c},
q_i^t
\right).
\end{equation}

\paragraph{Unavailable-modality-state anchoring.}
For unavailable modalities, CSA predicts teacher-side modality states instead of reconstructing raw inputs, providing modality-level references without requiring raw missing-input recovery.
Let \(\mathcal R_i^c=\mathcal M\setminus\mathcal A_i^c\) be the unavailable modality set, \(\mathbf a_i^c\) the modality-availability indicator, and \(\mathbf z_i^c=[\boldsymbol\mu_i^{s,c};\boldsymbol\sigma_i^{s,c};\mathbf a_i^c]\) the decoder input.
For each \(m\in\mathcal R_i^c\), a lightweight state decoder predicts the corresponding teacher-side Gaussian-inspired state:
\begin{equation}
\begin{aligned}
\widehat q_{m,i}^{s,c}
&=
D_m
\left(
\mathbf z_i^c
\right),\\
\widehat q_{m,i}^{s,c}
&=
\mathcal G
\left(
\widehat{\boldsymbol\mu}_{m,i}^{s,c},
\widehat{\boldsymbol\sigma}_{m,i}^{s,c}
\right).
\end{aligned}
\end{equation}
The decoder follows the same Gaussian-inspired state parameterization as \(\operatorname{StateHead}_{m}\), but uses independent parameters.

For compactness, define the unavailable-modality state distance as
\(d_{m,i}^{c}=\mathcal D_G(\widehat q_{m,i}^{s,c},q_{m,i}^{t})\).
The unavailable-modality-state anchoring loss is
\begin{equation}
\mathcal L_{\mathrm{mstate}}^{i,c}
=
\begin{cases}
\displaystyle
\frac{1}{|\mathcal R_i^c|}
\sum_{m\in\mathcal R_i^c}
d_{m,i}^{c},
&
|\mathcal R_i^c|>0,\\[0.8ex]
0,
&
|\mathcal R_i^c|=0 .
\end{cases}
\end{equation}

The CSA training objective is
\begin{equation}
\begin{aligned}
\mathcal L_{\mathrm{CSA}}
&=
\mathbb E_{i,c}
\Big[
\mathcal L_{\mathrm{pred}}^{i,c}
+
\lambda_{\mathrm{state}}
\mathcal L_{\mathrm{state}}^{i,c}\\
&\qquad\quad+
\lambda_{\mathrm{mstate}}
\mathcal L_{\mathrm{mstate}}^{i,c}
\Big],
\end{aligned}
\end{equation}
where the condition \(c\) is sampled from the training observation conditions.

\subsection{Nonverbal Conflict Exposure}
\label{sec:NCE}

NCE builds target-preserving conflict views as controlled perturbations by keeping the target language and label unchanged while replacing audio/visual cues with different-label donor cues, thereby exposing the student to misleading nonverbal observations while discouraging reliance on target-inconsistent nonverbal cues.

\paragraph{Conflict-view construction.}
For target sample \(i\), a donor is a mini-batch sample with a different label:
\begin{equation}
\mathcal J_i
=
\{j\in\mathcal I_{\mathrm{mb}}\mid y_j\neq y_i\}.
\end{equation}
If \(\mathcal J_i=\emptyset\), sample \(i\) is skipped for NCE; otherwise, \(j\sim\mathrm{Unif}(\mathcal J_i)\).
Given donor \(j\), we sample a nonverbal replacement subset
\(b\sim\mathrm{Unif}(\mathcal B)\), where
\(\mathcal B=\{\{\mathrm a\},\{\mathrm v\},\{\mathrm a,\mathrm v\}\}\), and construct
\begin{equation}
\widetilde{\mathbf x}_i^{m,b}
=
\begin{cases}
\mathbf x_j^{m,\mathrm{full}}, & m\in b,\\
\mathbf x_i^{m,\mathrm{full}}, & m\notin b,
\end{cases}
\qquad
m\in\mathcal M .
\end{equation}
Because the language input and target-side supervision are kept fixed by construction, we use \(y_i\) as the conflict-view training label.
The donor label \(y_j\) is used only for donor selection and margin interpretation.
The student processes
\(\widetilde{\mathbf X}_i^b
=
\{\widetilde{\mathbf x}_i^{m,b}\}_{m\in\mathcal M}\)
and outputs \(\widetilde{\mathbf o}_i^{s,b}\).

\paragraph{Margin interpretation and NCE loss.}
Let
\(\Delta_{ij}^{b}
=
\widetilde{\mathbf o}_{i,y_i}^{s,b}
-
\widetilde{\mathbf o}_{i,y_j}^{s,b}\)
be the target-versus-donor margin.
Since \(y_j\neq y_i\), the donor label is a non-target class in cross-entropy:
\begingroup
\small
\begin{equation}
\mathrm{CrossEntropy}
\left(
y_i,\widetilde{\mathbf o}_i^{s,b}
\right)
\ge
\log\left(1+\exp(-\Delta_{ij}^{b})\right).
\end{equation}
\endgroup
Thus, target-label cross-entropy implicitly discourages donor-label bias.
The NCE loss is
\begin{equation}
\mathcal L_{\mathrm{NCE}}
=
\mathbb E_{i,b,j}
\left[
\mathrm{CrossEntropy}\left(y_i,\widetilde{\mathbf o}_i^{s,b}\right)
\right],
\end{equation}
where \(i\) is valid, \(b\sim\mathrm{Unif}(\mathcal B)\), and \(j\sim\mathrm{Unif}(\mathcal J_i)\).
NCE is used only during training.

\paragraph{Final objective.}
The objective of CoRe-KD is
\begin{equation}
\mathcal L
=
\mathcal L_{\mathrm{CSA}}
+
\lambda_{\mathrm{NCE}}
\mathcal L_{\mathrm{NCE}}.
\end{equation}
During training, the teacher remains fixed.
At inference time, only the student encoders, state heads, fusion module, and classifier are retained.

\begin{table*}[t]
\centering
\small
\setlength{\tabcolsep}{2.4pt}
\renewcommand{\arraystretch}{0.98}
\setlength{\arrayrulewidth}{0.4pt}
\setlength{\doublerulesep}{1.2pt}

\newcommand{\oursc}{\cellcolor{gray!12}}
\newcommand{\gainc}{\cellcolor{gray!6}}
\providecommand{\best}[1]{\textbf{#1}}
\providecommand{\second}[1]{\underline{#1}}
\providecommand{\ph}{xx.xx/xx.xx}

\resizebox{0.99\textwidth}{!}{%
\begin{tabular}{@{}c|l||*{7}{c}@{}}
\Xhline{0.9pt}
\multicolumn{2}{c||}{Testing Condition}
& $\{\mathrm a\}$ & $\{\mathrm v\}$ & $\{\mathrm \ell\}$
& $\{\mathrm \ell,\mathrm a\}$ & $\{\mathrm a,\mathrm v\}$
& $\{\mathrm \ell,\mathrm v\}$ & $\{\mathrm \ell,\mathrm a,\mathrm v\}$ \\
\hline
Dataset & Method
& Acc(\%)/F1(\%) & Acc(\%)/F1(\%) & Acc(\%)/F1(\%)
& Acc(\%)/F1(\%) & Acc(\%)/F1(\%) & Acc(\%)/F1(\%)
& Acc(\%)/F1(\%) \\
\hline

\multirow{8}{*}{\begin{tabular}{c}IEMOCAP\\Six\end{tabular}}
& IMDer
& 47.13/46.58 & 36.78/29.35 & 65.56/65.58
& 67.41/67.23 & 50.22/49.44 & 67.10/67.00 & 68.15/68.12 \\
& Corr-KD
& 52.37/51.17 & 28.03/22.75 & 65.56/65.59
& \second{72.64}/\second{72.72} & 57.30/56.61 & 67.28/67.31 & \second{72.52}/\second{72.64} \\
& LNLN
& 47.69/45.71 & 37.89/\second{34.12} & 64.63/64.55
& 67.96/67.92 & 54.04/53.79 & 65.99/66.01 & 69.13/69.12 \\
& MoMKE
& 48.80/47.06 & 35.24/29.84 & 65.19/65.36
& 68.70/68.79 & 51.88/51.45 & 67.53/\second{67.68} & 69.93/70.06 \\
& MCULoRA
& \second{53.17}/\second{51.26} & 38.32/30.82 & \second{65.93}/65.94
& 69.25/69.23 & \second{60.32}/\second{59.70} & \second{67.65}/67.15 & 70.92/70.85 \\
& ComP
& 52.87/50.17 & \second{39.68}/30.43 & 65.87/\second{65.96}
& 69.19/69.03 & 58.78/57.47 & 67.28/67.04 & 71.04/70.95 \\
& \oursc CoRe-KD
& \oursc \best{57.86}/\best{57.82} & \oursc \best{40.36}/\best{36.36} & \oursc \best{67.53}/\best{67.56}
& \oursc \best{73.75}/\best{73.83} & \oursc \best{61.12}/\best{61.13} & \oursc \best{68.58}/\best{68.53} & \oursc \best{74.68}/\best{74.74} \\
& \gainc $\Delta$SOTA
& \gainc $\uparrow$ 4.69/6.56 & \gainc $\uparrow$ 0.68/2.24 & \gainc $\uparrow$ 1.60/1.60
& \gainc $\uparrow$ 1.11/1.11 & \gainc $\uparrow$ 0.80/1.43 & \gainc $\uparrow$ 0.93/0.85 & \gainc $\uparrow$ 2.16/2.10 \\
\hline

\multirow{8}{*}{\begin{tabular}{c}MELD\\Seven\end{tabular}}
& IMDer
& 35.03/20.41 & 35.03/20.41 & \second{66.54}/65.36
& 66.54/65.37 & 35.03/20.41 & \second{66.53}/65.36 & 66.54/65.36 \\
& Corr-KD
& 44.87/33.23 & 41.45/30.38 & 65.56/\second{65.39}
& 66.00/\second{65.78} & 43.69/33.39 & 65.61/\second{65.49} & 65.95/65.81 \\
& LNLN
& 48.28/\second{36.88} & 48.04/\second{31.50} & 66.26/65.21
& 66.62/65.75 & 48.64/\second{38.14} & 66.25/65.27 & 66.69/\second{65.86} \\
& MoMKE
& 48.07/32.72 & \second{48.12}/31.27 & 63.87/64.50
& 66.00/65.58 & 48.37/31.93 & 65.91/65.47 & \second{66.96}/65.68 \\
& MCULoRA
& 48.11/34.93 & 48.04/31.45 & 65.96/64.93
& 66.08/65.29 & 49.04/36.30 & 65.93/64.76 & 66.19/65.26 \\
& ComP
& \second{49.23}/36.42 & \second{48.12}/31.27 & 66.07/64.57
& \second{66.73}/65.35 & \second{49.25}/37.15 & 66.05/64.62 & 66.74/65.44 \\
& \oursc CoRe-KD
& \oursc \best{49.77}/\best{41.08} & \oursc \best{48.35}/\best{32.24} & \oursc \best{68.28}/\best{67.25}
& \oursc \best{68.47}/\best{67.35} & \oursc \best{50.19}/\best{38.59} & \oursc \best{68.66}/\best{67.97} & \oursc \best{68.85}/\best{67.41} \\
& \gainc $\Delta$SOTA
& \gainc $\uparrow$ 0.54/4.20 & \gainc $\uparrow$ 0.23/0.74 & \gainc $\uparrow$ 1.74/1.86
& \gainc $\uparrow$ 1.74/1.57 & \gainc $\uparrow$ 0.94/0.46 & \gainc $\uparrow$ 2.13/2.48 & \gainc $\uparrow$ 1.89/1.55 \\
\hline

\multicolumn{2}{c||}{Missing Rate}
& 0.1 & 0.2 & 0.3 & 0.4 & 0.5 & 0.6 & 0.7 \\
\hline
Dataset & Method
& Acc(\%)/F1(\%) & Acc(\%)/F1(\%) & Acc(\%)/F1(\%)
& Acc(\%)/F1(\%) & Acc(\%)/F1(\%) & Acc(\%)/F1(\%)
& Acc(\%)/F1(\%) \\
\hline

\multirow{8}{*}{\begin{tabular}{c}IEMOCAP\\Six\end{tabular}}
& IMDer
& 67.10/66.99 & 65.80/65.55 & 64.82/64.71
& 62.66/62.45 & 59.52/59.39 & 57.98/57.70 & 54.53/54.09 \\
& Corr-KD
& 69.99/70.01 & 68.21/68.24 & 67.10/67.17
& 65.31/65.42 & 63.46/63.07 & 61.18/60.64 & 54.47/53.65 \\
& LNLN
& 68.21/68.26 & 67.22/67.16 & 65.13/65.31
& 63.65/63.50 & 61.31/61.02 & 60.26/59.86 & 54.78/54.44 \\
& MoMKE
& 69.38/69.56 & 66.17/66.33 & 64.82/65.05
& 63.34/63.58 & 60.20/60.35 & 58.90/58.06 & 55.33/54.23 \\
& MCULoRA
& 70.18/70.08 & \second{68.76}/\second{68.63} & \second{67.53}/\second{67.32}
& \second{66.05}/\second{65.77} & 63.15/62.90 & 61.49/61.13 & 57.79/\second{57.93} \\
& ComP
& \second{70.24}/\second{70.18} & 68.45/68.02 & 67.47/66.83
& 65.68/65.35 & \second{64.39}/\second{64.20} & \second{61.98}/\second{61.84} & \second{57.98}/56.42 \\
& \oursc CoRe-KD
& \oursc \best{73.55}/\best{73.53} & \oursc \best{72.83}/\best{72.88} & \oursc \best{71.37}/\best{71.49}
& \oursc \best{71.35}/\best{71.59} & \oursc \best{70.12}/\best{70.19} & \oursc \best{69.25}/\best{69.29} & \oursc \best{68.45}/\best{68.30} \\
& \gainc $\Delta$SOTA
& \gainc $\uparrow$ 3.31/3.35 & \gainc $\uparrow$ 4.07/4.25 & \gainc $\uparrow$ 3.84/4.17
& \gainc $\uparrow$ 5.30/5.82 & \gainc $\uparrow$ 5.73/5.99 & \gainc $\uparrow$ 7.27/7.45 & \gainc $\uparrow$ 10.47/10.37 \\
\hline

\multirow{8}{*}{\begin{tabular}{c}MELD\\Seven\end{tabular}}
& IMDer
& 64.41/62.73 & 62.63/60.41 & 60.92/57.97
& 58.96/55.09 & 57.23/52.32 & 55.33/49.12 & 53.43/45.32 \\
& Corr-KD
& 64.09/\second{63.77} & 61.85/\second{61.36} & 59.31/58.51
& 56.58/55.48 & 54.28/52.76 & 52.08/50.15 & 49.23/\second{46.73} \\
& LNLN
& 64.80/63.65 & 62.67/61.12 & 60.98/\second{58.81}
& 59.08/\second{56.10} & 57.66/\second{53.78} & 55.61/\second{50.48} & 53.52/46.39 \\
& MoMKE
& \second{65.12}/63.29 & \second{63.12}/60.69 & 61.21/58.06
& 59.64/55.58 & 57.93/52.66 & 55.71/49.03 & 53.92/45.52 \\
& MCULoRA
& 64.49/63.05 & 62.86/60.78 & 61.26/58.45
& 59.44/55.61 & 57.80/52.90 & 55.98/49.74 & 53.84/45.60 \\
& ComP
& 64.76/62.99 & 63.01/60.69 & \second{61.49}/58.50
& \second{59.75}/55.87 & \second{58.01}/53.15 & \second{56.18}/50.04 & \second{54.07}/46.07 \\
& \oursc CoRe-KD
& \oursc \best{66.36}/\best{65.56} & \oursc \best{64.18}/\best{63.28} & \oursc \best{61.84}/\best{60.70}
& \oursc \best{60.11}/\best{57.61} & \oursc \best{58.81}/\best{56.55} & \oursc \best{56.32}/\best{52.78} & \oursc \best{54.18}/\best{48.42} \\
& \gainc $\Delta$SOTA
& \gainc $\uparrow$ 1.24/1.79 & \gainc $\uparrow$ 1.06/1.93 & \gainc $\uparrow$ 0.34/1.89
& \gainc $\uparrow$ 0.37/1.51 & \gainc $\uparrow$ 0.80/2.76 & \gainc $\uparrow$ 0.14/2.30 & \gainc $\uparrow$ 0.11/1.69 \\
\Xhline{0.9pt}
\end{tabular}%
}
\caption{Results under fixed-missing and random-missing settings on IEMOCAP-6 and MELD-7.}
\label{tab:missing_results_iemocap6_meld7}
\end{table*}

\section{Experiment}
\subsection{Dataset and Evaluation Metrics}
\paragraph{Datasets.}
We evaluate CoRe-KD on IEMOCAP~\citep{busso2008iemocap} under 4-/6-class protocols and MELD~\citep{poria2019meld} under the standard 7-class protocol as the primary dialogue-context MER benchmarks. We further report CMU-MOSEI~\cite{bagher-zadeh-etal-2018-multimodal} results in Table~\ref{tab:mosei_generalization}
as a supplementary utterance-level generalization check, since MOSEI does not
evaluate dialogue-context modeling.

\paragraph{Evaluation Protocols and Metrics.}
In prior IEMOCAP-based MER studies, accuracy-based metrics are commonly reported.
However, MELD-7 has under-represented classes such as fear and disgust.
We therefore report accuracy (Acc) and weighted-F1 (F1) for all evaluations, with F1 as the primary metric. 
We evaluate all methods under two missing-modality protocols.

\noindent\textit{Fixed Missing Protocol} follows MMIN~\citep{zhao2021mmin} and evaluates fixed available subsets from
\(\{a\}\), \(\{v\}\), \(\{l\}\), \(\{l,a\}\), \(\{a,v\}\), \(\{l,v\}\), and \(\{l,a,v\}\).
The same subset is applied to all test samples, and results are reported per condition.

\noindent\textit{Random Missing Protocol} follows IMDer~\citep{wang2023imder} and samples per-sample missing patterns with ratios from 0.1 to 0.7, while retaining at least one modality available. All results are reported per missing ratio.

\begin{table*}[t]
\centering
\small
\setlength{\tabcolsep}{3.2pt}
\renewcommand{\arraystretch}{1.04}
\setlength{\arrayrulewidth}{0.4pt}
\setlength{\doublerulesep}{1.2pt}

\providecommand{\oursc}{\cellcolor{gray!12}}
\providecommand{\best}[1]{\textbf{#1}}

\resizebox{0.99\textwidth}{!}{%
\begin{tabular}{@{}c|l||*{7}{c}@{}}
\Xhline{0.9pt}
\multicolumn{2}{c||}{Testing Condition}
& $\{\mathrm a\}$ & $\{\mathrm v\}$ & $\{\mathrm \ell\}$ 
& $\{\mathrm \ell,\mathrm a\}$ & $\{\mathrm a,\mathrm v\}$ 
& $\{\mathrm \ell,\mathrm v\}$ & $\{\mathrm \ell,\mathrm a,\mathrm v\}$ \\
\hline
Dataset & Method
& Acc(\%)/F1(\%) & Acc(\%)/F1(\%) & Acc(\%)/F1(\%)
& Acc(\%)/F1(\%) & Acc(\%)/F1(\%) & Acc(\%)/F1(\%)
& Acc(\%)/F1(\%) \\
\hline

\multirow{7}{*}{CMU-MOSEI}
& IMDer
& 68.55/67.14 & 63.07/63.27 & 85.55/85.41
& 85.25/85.04 & 67.36/67.16 & 84.78/84.67 & 85.22/85.10 \\
& Corr-KD
& 63.79/51.12 & 66.57/61.41 & 84.37/84.37
& 84.73/84.73 & 68.38/64.76 & 83.13/83.17 & 82.72/82.76 \\
& LNLN
& 71.46/69.01 & 65.30/60.34 & 86.19/86.26
& 86.54/86.61 & 71.66/69.96 & 86.19/86.24 & 86.16/86.20 \\
& MoMKE
& 70.53/68.56 & 66.57/61.97 & 85.44/85.61
& 86.16/86.24 & 71.60/68.68 & 86.13/86.18 & 86.05/85.98 \\
& MCULoRA
& 68.93/65.75 & 65.60/60.55 & 85.97/86.01
& 86.54/86.55 & 70.47/68.40 & 86.16/86.14 & 86.05/85.98 \\
& ComP
& 69.07/67.50 & 66.29/62.67 & 86.19/86.21
& 86.02/86.02 & 71.05/69.69 & \best{86.82}/86.77 & 85.86/85.80 \\
& \oursc CoRe-KD
& \oursc \best{72.95}/\best{72.39} & \oursc \best{67.01}/\best{67.24} & \oursc \best{86.49}/\best{86.49}
& \oursc \best{86.98}/\best{86.98} & \oursc \best{72.81}/\best{72.67} & \oursc \best{86.82}/\best{86.85} & \oursc \best{87.12}/\best{87.16} \\
\hline

\multicolumn{2}{c||}{Missing Rate}
& 0.1 & 0.2 & 0.3 & 0.4 & 0.5 & 0.6 & 0.7 \\
\hline
Dataset & Method
& Acc(\%)/F1(\%) & Acc(\%)/F1(\%) & Acc(\%)/F1(\%)
& Acc(\%)/F1(\%) & Acc(\%)/F1(\%) & Acc(\%)/F1(\%)
& Acc(\%)/F1(\%) \\
\hline

\multirow{7}{*}{CMU-MOSEI}
& IMDer
& 85.88/85.93 & 84.18/84.17 & 83.49/83.41
& 82.55/82.36 & 79.86/79.37 & 77.90/77.01 & 74.99/73.23 \\
& Corr-KD
& 82.61/82.34 & 81.73/81.26 & 80.30/79.46
& 77.49/75.96 & 74.77/72.16 & 71.85/67.63 & 69.21/63.17 \\
& LNLN
& 85.75/85.66 & 84.48/84.33 & 82.94/82.63
& 81.07/80.48 & 79.33/78.35 & 76.66/74.93 & 74.22/71.55 \\
& MoMKE
& 86.24/86.21 & 84.73/84.64 & 83.43/83.19
& 81.43/80.91 & 79.36/78.38 & 76.33/74.52 & 72.81/69.49 \\
& MCULoRA
& 85.72/85.71 & 84.67/84.60 & 83.49/83.33
& 82.09/81.80 & 80.77/80.29 & 78.65/77.72 & 76.58/74.76 \\
& ComP
& 85.70/85.65 & 84.86/84.72 & 84.05/83.88
& 83.32/83.05 & 81.82/81.35 & \best{81.28}/\best{80.58} & \best{79.18}/\best{78.02} \\
& \oursc CoRe-KD
& \oursc \best{86.49}/\best{86.40} & \oursc \best{85.06}/\best{84.89} & \oursc \best{84.23}/\best{84.04}
& \oursc \best{83.52}/\best{83.21} & \oursc \best{82.00}/\best{81.54} & \oursc 81.10/80.42 & \oursc {78.92}/{77.77} \\
\Xhline{0.9pt}
\end{tabular}%
}
\caption{Supplementary generalization results on CMU-MOSEI under fixed-missing and random-missing settings. MOSEI is used as an utterance-level generalization check rather than an exhaustive SOTA comparison.}
\label{tab:mosei_generalization}
\end{table*}

\subsection{Implementation Details}

\paragraph{CoRe-KD implementation.}
For CoRe-KD, the complete-view teacher is first trained on clean training data and then frozen.
The student is trained with fixed-missing, random-missing, and NCE auxiliary views, without an additional clean-view-only student training stage.
For random-missing views, the missing ratio is sampled from \(0.1\) to \(0.7\), while retaining at least one modality.
We report mean performance over five random seeds under the same fixed- and random-missing evaluation protocols.
Hyperparameters are provided in Appendix~\ref{app:training_details}.

\paragraph{Baseline implementation.}
For distillation baselines, we use the same clean complete-view teacher as CoRe-KD whenever applicable.
For reproduced baselines, we use the same language, acoustic, and visual features as CoRe-KD whenever supported, and train/evaluate them under the same missing-modality protocols and seed settings.
All experiments are run on four NVIDIA RTX A5000 GPUs. Detailed controls on data splits, feature inputs, checkpoint selection, and evaluation masks are provided in Appendix~\ref{app:baseline}.

\begin{table*}[t]
\centering
\small
\setlength{\tabcolsep}{3.2pt}
\renewcommand{\arraystretch}{1.04}
\setlength{\arrayrulewidth}{0.4pt}
\setlength{\doublerulesep}{1.2pt}

\providecommand{\abph}{xx.xx/xx.xx}
\providecommand{\abours}{\cellcolor{gray!12}}
\providecommand{\best}[1]{\textbf{#1}}

\resizebox{0.99\textwidth}{!}{%
\begin{tabular}{@{}c|l||*{7}{c}@{}}
\Xhline{0.9pt}
\multicolumn{2}{c||}{Testing Condition}
& $\{\mathrm a\}$ & $\{\mathrm v\}$ & $\{\mathrm \ell\}$
& $\{\mathrm \ell,\mathrm a\}$ & $\{\mathrm a,\mathrm v\}$
& $\{\mathrm \ell,\mathrm v\}$ & $\{\mathrm \ell,\mathrm a,\mathrm v\}$ \\
\hline
Dataset & Variant
& Acc(\%)/F1(\%) & Acc(\%)/F1(\%) & Acc(\%)/F1(\%)
& Acc(\%)/F1(\%) & Acc(\%)/F1(\%) & Acc(\%)/F1(\%)
& Acc(\%)/F1(\%) \\
\hline

\multirow{6}{*}{\begin{tabular}{c}IEMOCAP\\Six\end{tabular}}
& w/o $\mathcal{L}_{\mathrm{CSA}}$
& 46.27/45.29 & 33.33/32.45 & 63.52/63.61
& 72.15/72.41 & 50.28/48.92 & 65.43/65.66 & 72.89/73.12 \\

& w/o $\mathcal{L}_{\mathrm{NCE}}$
& 53.60/53.63 & 37.28/\best{36.73} & 65.37/65.24
& 72.52/72.54 & 58.90/58.90 & 66.67/66.72 & 73.81/73.88 \\

& w/o $\mathcal{L}_{\mathrm{pred}}$
& 47.26/46.31 & 32.47/29.44 & 64.70/64.85
& 72.21/72.27 & 50.83/49.66 & 65.99/66.08 & 72.64/72.72 \\

& w/o $\mathcal{L}_{\mathrm{state}}$
& 54.90/54.15 & 36.29/32.29 & 67.22/67.15
& 73.26/73.22 & 58.90/58.14 & 68.21/68.13 & 73.88/73.83 \\

& w/o $\mathcal{L}_{\mathrm{mstate}}$
& 53.85/53.76 & 34.38/27.38 & 67.28/67.37
& 73.44/73.38 & 57.42/57.32 & 68.21/68.45 & 74.00/74.05 \\

& \abours CoRe-KD
& \abours \best{57.86}/\best{57.82} & \abours \best{40.36}/36.36 & \abours \best{67.53}/\best{67.56}
& \abours \best{73.75}/\best{73.83} & \abours \best{61.12}/\best{61.13} & \abours \best{68.58}/\best{68.53} & \abours \best{74.68}/\best{74.74} \\
\hline

\multicolumn{2}{c||}{Missing Rate}
& 0.1 & 0.2 & 0.3 & 0.4 & 0.5 & 0.6 & 0.7 \\
\hline
Dataset & Variant
& Acc(\%)/F1(\%) & Acc(\%)/F1(\%) & Acc(\%)/F1(\%)
& Acc(\%)/F1(\%) & Acc(\%)/F1(\%) & Acc(\%)/F1(\%)
& Acc(\%)/F1(\%) \\
\hline

\multirow{6}{*}{\begin{tabular}{c}IEMOCAP\\Six\end{tabular}}
& w/o $\mathcal{L}_{\mathrm{CSA}}$
& 70.67/70.91 & 70.67/70.89 & 69.56/69.78
& 68.88/69.05 & 66.97/67.31 & 66.36/66.79 & 63.96/64.31 \\

& w/o $\mathcal{L}_{\mathrm{NCE}}$
& 72.15/72.21 & 71.10/71.20 & 69.81/69.92
& 68.58/68.71 & 68.02/68.27 & 67.90/68.21 & 66.48/66.70 \\

& w/o $\mathcal{L}_{\mathrm{pred}}$
& 70.43/70.52 & 70.30/70.38 & 69.99/70.07
& 69.44/69.39 & 68.76/68.90 & 66.91/67.08 & 63.83/63.71 \\

& w/o $\mathcal{L}_{\mathrm{state}}$
& 72.58/72.59 & 71.10/71.10 & 70.61/70.67
& 69.99/70.03 & 69.13/69.15 & 68.82/68.86 & 65.93/65.76 \\

& w/o $\mathcal{L}_{\mathrm{mstate}}$
& 71.78/71.88 & 70.67/70.77 & 70.12/70.23
& 70.61/70.75 & 69.62/69.80 & 68.64/68.61 & 66.30/66.21 \\

& \abours CoRe-KD
& \abours \best{73.55}/\best{73.53} & \abours \best{72.83}/\best{72.88} & \abours \best{71.37}/\best{71.49}
& \abours \best{71.35}/\best{71.59} & \abours \best{70.12}/\best{70.19} & \abours \best{69.25}/\best{69.29} & \abours \best{68.45}/\best{68.30} \\
\Xhline{0.9pt}
\end{tabular}%
}
\caption{Ablation of CoRe-KD on IEMOCAP Six under fixed-missing testing conditions and random-missing rates.}
\label{tab:ablation_iemocap6}
\end{table*}

\begin{table*}[t]
\centering
\small
\setlength{\tabcolsep}{3.2pt}
\renewcommand{\arraystretch}{1.04}
\setlength{\arrayrulewidth}{0.4pt}
\setlength{\doublerulesep}{1.2pt}

\newcommand{\oursc}{\cellcolor{gray!12}}
\newcommand{\gainc}{\cellcolor{gray!6}}

\resizebox{0.99\textwidth}{!}{%
\begin{tabular}{@{}c|l||*{7}{c}@{}}
\Xhline{0.9pt}
\multicolumn{2}{c||}{Testing Condition}
& $\{a\}$ & $\{v\}$ & $\{l\}$ & $\{l,a\}$ & $\{a,v\}$ & $\{l,v\}$ & $\{l,a,v\}$ \\
\hline
Dataset & Method
& Acc(\%)/F1(\%) & Acc(\%)/F1(\%) & Acc(\%)/F1(\%)
& Acc(\%)/F1(\%) & Acc(\%)/F1(\%) & Acc(\%)/F1(\%)
& Acc(\%)/F1(\%) \\
\hline

\multirow{5}{*}{\shortstack{IEMOCAP\\Four}}
& MulT + Vanilla KD
& 45.97/43.39 & 41.71/31.05 & 76.73/76.74 & 77.62/77.78 & 56.84/56.99 & 74.07/73.79 & 73.51/73.54 \\
& MulT + DKD
& 52.42/49.87 & 45.41/37.59 & 69.57/69.13 & 77.62/77.74 & 57.41/56.81 & 70.13/69.72 & 76.25/76.33 \\
& MulT + Corr-KD
& 52.33/49.62 & 47.18/38.78 & 79.63/79.62 & 80.43/80.59 & 65.94/65.89 & 78.02/78.10 & 76.65/76.81 \\
& \oursc MulT + CoRe-KD
& \oursc \textbf{54.91/50.25} & \oursc \textbf{47.91/42.29} & \oursc \textbf{79.79/79.95}
& \oursc \textbf{80.52/80.76} & \oursc \textbf{72.06/72.19} & \oursc \textbf{81.40/81.42}
& \oursc \textbf{82.77/82.94} \\
& \gainc $\Delta$
& \gainc $+2.49/+0.38$ & \gainc $+0.73/+3.51$ & \gainc $+0.16/+0.33$
& \gainc $+0.09/+0.17$ & \gainc $+6.12/+6.30$ & \gainc $+3.38/+3.32$
& \gainc $+6.12/+6.13$ \\
\hline

\multicolumn{2}{c||}{Testing Condition}
& $\{a\}$ & $\{v\}$ & $\{l\}$ & $\{l,a\}$ & $\{a,v\}$ & $\{l,v\}$ & $\{l,a,v\}$ \\
\hline
Dataset & Method
& Acc(\%)/F1(\%) & Acc(\%)/F1(\%) & Acc(\%)/F1(\%)
& Acc(\%)/F1(\%) & Acc(\%)/F1(\%) & Acc(\%)/F1(\%)
& Acc(\%)/F1(\%) \\
\hline

\multirow{5}{*}{\shortstack{IEMOCAP\\Four}}
& MISA + Vanilla KD
& 44.61/37.79 & 44.85/37.00 & 63.20/61.68 & 71.42/71.40 & 48.23/43.39 & 63.28/59.84 & 71.42/71.57 \\
& MISA + DKD
& 45.81/41.82 & 40.34/31.47 & 64.33/63.67 & 72.30/72.45 & 54.59/54.66 & 64.17/62.88 & 75.36/75.42 \\
& MISA + Corr-KD
& 52.74/52.81 & 50.16/40.24 & 66.67/66.48 & 73.67/73.69 & 60.71/60.94 & 68.28/68.12 & 75.52/75.52 \\
& \oursc MISA + CoRe-KD
& \oursc \textbf{54.83/53.78} & \oursc \textbf{52.98/50.08} & \oursc \textbf{67.87/68.06}
& \oursc \textbf{75.12/75.27} & \oursc \textbf{63.53/63.66} & \oursc \textbf{73.51/73.54}
& \oursc \textbf{77.54/77.58} \\
& \gainc $\Delta$
& \gainc $+2.09/+0.97$ & \gainc $+2.82/+9.84$ & \gainc $+1.20/+1.58$
& \gainc $+1.45/+1.58$ & \gainc $+2.82/+2.72$ & \gainc $+5.23/+5.42$
& \gainc $+2.02/+2.06$ \\
\Xhline{0.9pt}
\end{tabular}%
}
\caption{
Teacher-backbone generalization of CoRe-KD on IEMOCAP Four under fixed-missing testing conditions.
$\Delta$ denotes the improvement of CoRe-KD over the strongest KD baseline under the same complete-view teacher.
}
\label{tab:teacher_generalization_iemocap4}
\end{table*}

\subsection{Baselines}
We compare CoRe-KD with representative robust multimodal emotion recognition methods under two missing-modality protocols: fixed missing and random missing.
The baselines include reconstruction- or recovery-based methods, i.e., IMDer~\citep{wang2023imder} and LNLN~\citep{zhang2024lnln}; distillation- or expert-knowledge-transfer-based methods, i.e., Corr-KD~\citep{li2024corrkd} and MoMKE~\citep{xu2024momke}; and adaptation- or prompting-based methods, i.e., MCULoRA~\citep{zhao2025mculora} and ComP~\citep{he2026comp}.
Together, these methods cover recent representative solutions for incomplete multimodal observations.

\subsection{Main Result}

\paragraph{Main results.}
Table~\ref{tab:missing_results_iemocap6_meld7} reports the results on IEMOCAP-6 and MELD-7.
CoRe-KD consistently improves missing-modality performance across fixed- and random-missing protocols.
It achieves the best Acc/F1 in all conditions on IEMOCAP-6, obtains the best Acc/F1 in all fixed-missing conditions on MELD-7, and also achieves the best Acc/F1 at all random-missing rates on MELD-7.
In average F1, CoRe-KD improves over the strongest baselines by 3.57/1.90 points under fixed missing and 6.22/2.08 points under random missing on IEMOCAP-6/MELD-7.
The improvements also hold in the complete-view condition \(\{\mathrm \ell,\mathrm a,\mathrm v\}\), suggesting that robust missing-view training does not come at the cost of full-modality performance within this protocol.

\begin{figure}[t]
    \centering
    \includegraphics[width=\linewidth]{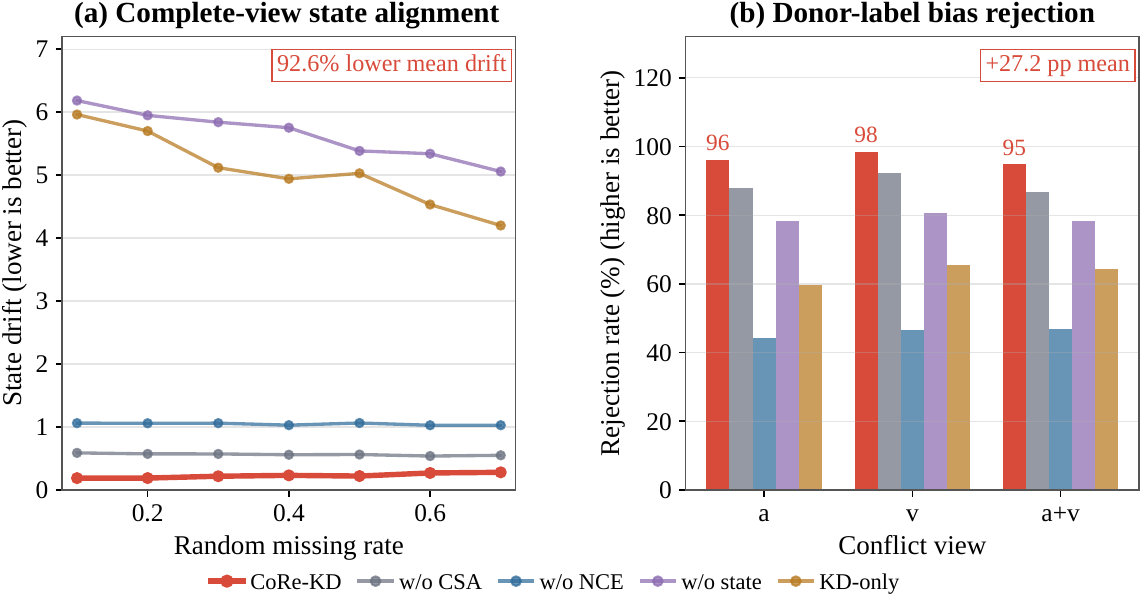}
    \caption{
    Mechanism analysis on IEMOCAP-6.
    (a) Lower state drift indicates better complete-view state alignment.
    (b) Higher rejection rate indicates better resistance to donor-label bias under nonverbal conflict.
    }
    \label{fig:mechanism}
    \vspace{-0.3cm}
\end{figure}

\paragraph{Supplementary generalization.}
Table~\ref{tab:mosei_generalization} reports CMU-MOSEI as a supplementary utterance-level generalization check.
CoRe-KD obtains the best Acc/F1 in fixed-missing conditions and at random-missing rates from 0.1 to 0.5, while staying close to the strongest baseline at 0.6 and 0.7.
In average F1, it gains 2.16 points under fixed missing and remains comparable under random missing, providing an additional check beyond the dialogue benchmarks.

\subsection{Ablation Study}
Table~\ref{tab:ablation_iemocap6} reports ablations on IEMOCAP-6.
CoRe-KD performs best.
Removing \(\mathcal{L}_{\mathrm{CSA}}\) or \(\mathcal{L}_{\mathrm{pred}}\) causes the largest drops, indicating that complete-view anchoring and prediction supervision are the main contributors; removing \(\mathcal{L}_{\mathrm{state}}\), \(\mathcal{L}_{\mathrm{mstate}}\), or \(\mathcal{L}_{\mathrm{NCE}}\) also hurts in most cases, suggesting complementary gains from fused-state alignment, unavailable-modality-state alignment, and nonverbal conflict regularization.
Appendix~\ref{app:controlled_state} controls for KD-only, hidden-state regression, raw reconstruction, deterministic state matching, and Gaussian CSA under the same backbone and training views, showing that the gains are not explained by prediction-level KD or generic auxiliary matching.

\section{Further Study}

To analyze the sources and teacher-backbone generality of CoRe-KD's gains, we conduct further studies around four questions.
Appendix~\ref{app:theory} gives lightweight theoretical notes on the design intuitions and scope.

\paragraph{Q1: Does CSA align with the complete-view teacher reference?}
We measure fused-state drift under the random-missing setting.
For condition \(c\), drift is
\(\mathbb{E}_{i}[(\|\mu^{s,c}_{i}-\mu^t_i\|^2_2+\|\sigma^{s,c}_{i}-\sigma^t_i\|^2_2)/d]\),
computed between the student fused state and the complete-view teacher state.
As shown in Figure~\ref{fig:mechanism}(a), CoRe-KD yields consistently lower drift than KD-only and ablated variants, with 92.6\% lower mean drift than the average of compared variants.
This indicates that CSA helps align the student state with the complete-view teacher reference beyond final prediction matching.

\begin{figure}[t]
    \centering
    \includegraphics[width=\linewidth]{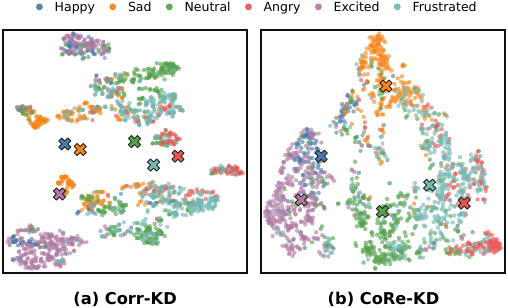}
    \caption{
    t-SNE visualization on IEMOCAP-6 under RMFM with missing rate 0.5.
    CoRe-KD yields more structured class distributions than Corr-KD.
    }
    \label{fig:tsne_iemocap6_rmfm05}
    \vspace{-0.3cm}
\end{figure}

\paragraph{Q2: Does NCE reduce donor-label bias?}
We measure rejection rates under target-preserving nonverbal conflict views.
For target \(i\), donor \(j\), and conflict type \(b\in\mathcal B\), rejection is counted when
\(d_\mu(\widetilde q^{s,b}_{i},q^t_i)<d_\mu(\widetilde q^{s,b}_{i},q^t_j)\),
where \(d_\mu(q,q')=\|\boldsymbol\mu(q)-\boldsymbol\mu(q')\|_2\).
As shown in Figure~\ref{fig:mechanism}(b), CoRe-KD improves the mean rejection rate from 69.3\% to 96.5\% across \(\{\mathrm a\}\), \(\{\mathrm v\}\), and \(\{\mathrm a,\mathrm v\}\) conflicts.
This suggests that NCE helps preserve target-side evidence under donor-label nonverbal cues.

\paragraph{Q3: Are the gains tied to a specific teacher?}
To avoid drawing teacher-specific conclusions, we evaluate CoRe-KD with two alternative complete-view teachers, MulT~\citep{tsai2019mult} and MISA~\citep{hazarika2020misa}, under the same fixed-missing protocol.
As shown in Table~\ref{tab:teacher_generalization_iemocap4}, CoRe-KD improves across both teacher backbones, suggesting that the benefit is not tied to a particular complete-view teacher architecture.

\paragraph{Q4: Does structured distillation help beyond prediction transfer?}
Within each teacher backbone, all variants share the same frozen teacher, student architecture, input features, and missing-view schedule, differing only in the transferred knowledge: prediction-level KD transfers logits~\citep{hinton2015distilling}, DKD decouples target/non-target knowledge~\citep{zhao2022decoupled}, Corr-KD transfers correlations, and CoRe-KD anchors fused and modality-specific Gaussian-inspired states.
The consistent gains under MulT and MISA suggest that structured complete-view state anchoring provides useful supervision beyond prediction-level transfer for incomplete-observation MER.

\section{Conclusion}
In this paper, we proposed CoRe-KD, a structured complete-view distillation framework for robust conversational multimodal emotion recognition under unreliable observations.
CoRe-KD uses Complete-view State Anchoring to align incomplete-view predictions and Gaussian-inspired states with complete-view teacher references, and Nonverbal Conflict Exposure to regularize target-preserving nonverbal conflict views.
Experiments on IEMOCAP and MELD show improved robustness under fixed- and random-missing protocols, with mechanism analyses suggesting better complete-view state alignment and target preservation under controlled nonverbal conflict, while retaining a single student model at inference.

\label{sec:conclusion}

\section*{Limitations}
CoRe-KD requires complete multimodal observations to train the frozen teacher, although inference only uses the student.
Its NCE module relies on controlled target-preserving conflict views and may not cover all real-world multimodal misalignment or corruption patterns.
In addition, the Gaussian-inspired state is used as a practical matching representation rather than a calibrated probabilistic posterior.
Future work may study CoRe-KD under more diverse missing, noisy, and naturally conflicting multimodal conditions.

\section*{Ethics Statement}
This work uses existing multimodal emotion recognition benchmarks and does not collect new human-subject data. For IEMOCAP, we obtained access through the official application procedure and used the data only for permitted academic research, without redistribution or any unauthorized use. MELD and CMU-MOSEI are publicly released research benchmarks and are used according to their intended research purposes. Since emotion recognition involves sensitive affective information, our results are intended for benchmark research only and should not be used for high-stakes decisions or user profiling without proper consent, privacy protection, and bias assessment.
\bibliography{custom}

\appendix

\counterwithin{equation}{section}
\counterwithin{table}{section}
\counterwithin{figure}{section}
\renewcommand{\theequation}{\thesection.\arabic{equation}}
\renewcommand{\thetable}{\thesection.\arabic{table}}
\renewcommand{\thefigure}{\thesection.\arabic{figure}}

\counterwithin{equation}{section}
\counterwithin{table}{section}
\counterwithin{figure}{section}
\renewcommand{\theequation}{\thesection.\arabic{equation}}
\renewcommand{\thetable}{\thesection.\arabic{table}}
\renewcommand{\thefigure}{\thesection.\arabic{figure}}

\appendix

\counterwithin{equation}{section}
\counterwithin{table}{section}
\counterwithin{figure}{section}
\renewcommand{\theequation}{\thesection.\arabic{equation}}
\renewcommand{\thetable}{\thesection.\arabic{table}}
\renewcommand{\thefigure}{\thesection.\arabic{figure}}

\appendix
\clearpage
\section{Experimental Protocol}
\label{app:reproducibility}

\subsection{Datasets}
\label{app:datasets}

\paragraph{IEMOCAP.}
IEMOCAP is used as one of our main dialogue-context MER benchmarks.
We report both the commonly used four-class and six-class protocols.
For IEMOCAP-4, we merge \textit{happy} and \textit{excited} and remove \textit{frustrated}, resulting in four classes:
\textit{happy/excited}, \textit{sad}, \textit{neutral}, and \textit{angry}.
For IEMOCAP-6, we keep six emotion classes:
\textit{happy}, \textit{sad}, \textit{neutral}, \textit{angry}, \textit{excited}, and \textit{frustrated}.

\paragraph{MELD.}
MELD is used as another main dialogue-context MER benchmark.
It contains multi-party conversations with seven emotion labels:
\textit{anger}, \textit{disgust}, \textit{fear}, \textit{joy}, \textit{neutral}, \textit{sadness}, and \textit{surprise}.
Compared with IEMOCAP, MELD provides more diverse conversational interactions and a more imbalanced label distribution.
In particular, \textit{fear} and \textit{disgust} are highly under-represented, making MELD a complementary testbed for evaluating robust conversational MER under imbalanced dialogue conditions.

\paragraph{CMU-MOSEI.}
CMU-MOSEI is included only as a supplementary utterance-level generalization check, rather than as a main dialogue-context benchmark.
Following prior robust multimodal sentiment analysis studies, we use the non-zero sentiment setting.
This setting helps examine whether CoRe-KD remains effective beyond dialogue-context MER, while the main claims are based on IEMOCAP and MELD. 

Detailed dataset splits and the final training split are summarized in Table~\ref{tab:dataset_splits}; all final models are trained on the official training split.

\subsection{Feature Extraction}
\label{app:features}

All experiments use pre-extracted utterance-level multimodal features.
The feature extractors are kept fixed, and CoRe-KD only trains the dialogue encoder, modality projection layers, state heads, fusion module, and classifier.
For dialogue-context datasets, utterance-level language, acoustic, and visual features are organized in dialogue order and processed as modality-specific sequences.
Missing-modality experiments are conducted by applying availability masks to these pre-extracted modality features, without changing the underlying feature extraction pipeline.

For IEMOCAP, textual features are obtained from a RoBERTa-Large encoder~\citep{liu2019roberta} fine-tuned on conversation transcripts, using the last-layer \texttt{[CLS]} representation.
Acoustic features are extracted with openSMILE~\citep{eyben2010opensmile}, and visual features are extracted with DenseNet~\citep{huang2017densely} from facial frames.
The resulting language, acoustic, and visual feature dimensions are \(1024\), \(1582\), and \(342\), respectively.

For MELD, textual features are extracted with a RoBERTa-based encoder~\citep{liu2019roberta} using speaker-aware dialogue context and a mask-prompt formulation, where the \texttt{<mask>} embedding is used as the utterance-level textual representation.
Acoustic features are extracted from the target utterance speech segment with data2vec~\citep{baevski2022data2vec}, and visual features are extracted from target utterance video frames with TimeSformer~\citep{bertasius2021space}.
In the processed MELD features, all three modalities are stored as \(768\)-dimensional utterance-level representations.

For CMU-MOSEI, audio, text, and visual embeddings are extracted with Wav2Vec~\citep{schneider2019wav2vec}, DeBERTa~\citep{he2021deberta}, and MA-Net~\citep{zhao2021learning}, respectively.
The resulting language, acoustic, and visual feature dimensions are \(1024\), \(512\), and \(1024\), respectively.
Each modality feature is first mapped to the model hidden dimension by a modality-specific projection layer before being passed to the intra-/inter-modal encoder.

\begin{table}[t]
\centering
\scriptsize
\setlength{\tabcolsep}{3.0pt}
\renewcommand{\arraystretch}{1.06}
\resizebox{\linewidth}{!}{%
\begin{tabular}{@{}l|c|c|c@{}}
\toprule
Dataset & Dialogues & Utterances & Final Train \\
\midrule
IEMOCAP-4
& 120 / -- / 31
& 4,342 / -- / 1,242
& Train \\

IEMOCAP-6
& 120 / -- / 31
& 5,810 / -- / 1,623
& Train \\

MELD-7
& 1,039 / 114 / 280
& 9,989 / 1,109 / 2,610
& Train \\

CMU-MOSEI
& -- / -- / --
& 12,786 / 1,438 / 3,634
& Train \\
\bottomrule
\end{tabular}%
}
\caption{
Dataset splits used in our experiments.
Each split is reported as Train / Dev / Test, and the Final Train column indicates the split used for final model training.
IEMOCAP and MELD are dialogue-context MER benchmarks; CMU-MOSEI is an utterance-level supplementary benchmark.
}
\label{tab:dataset_splits}
\end{table}
\begin{table}[t]
\centering
\scriptsize
\setlength{\tabcolsep}{4.0pt}
\renewcommand{\arraystretch}{1.05}
\resizebox{\linewidth}{!}{%
\begin{tabular}{@{}l|ccc|c@{}}
\toprule
Dataset & Language & Acoustic & Visual & Feature Level \\
\midrule
IEMOCAP & 1024 & 1582 & 342 & Utterance \\
MELD & 768 & 768 & 768 & Utterance \\
CMU-MOSEI & 1024 & 512 & 1024 & Utterance \\
\bottomrule
\end{tabular}%
}
\caption{
Input feature dimensions used in our implementation.
All features are pre-extracted and kept fixed during CoRe-KD training.
}
\label{tab:feature_dimensions}
\end{table}
\subsection{Missing-Modality Evaluation Protocols}
\label{app:missing_protocols}
\paragraph{Fixed missing.}
The fixed-missing protocol evaluates robustness under deterministic modality availability.
We consider seven available-modality subsets:
\(\{\mathrm a\}\), \(\{\mathrm v\}\), \(\{\mathrm \ell\}\),
\(\{\mathrm \ell,\mathrm a\}\), \(\{\mathrm a,\mathrm v\}\),
\(\{\mathrm \ell,\mathrm v\}\), and \(\{\mathrm \ell,\mathrm a,\mathrm v\}\),
where \(\mathrm \ell\), \(\mathrm a\), and \(\mathrm v\) denote language, acoustic, and visual modalities.
For each condition, the same modality subset is retained for all test samples, while the remaining modalities are treated as unavailable by the model.
The complete-view condition \(\{\mathrm \ell,\mathrm a,\mathrm v\}\) is included to check whether robust training preserves full-modality performance rather than only improving incomplete-view cases.

\paragraph{Random missing.}
The random-missing protocol evaluates sample-wise stochastic modality loss.
For each missing rate \(r\in\{0.1,0.2,\ldots,0.7\}\), we independently generate modality-availability masks for test samples and ensure that at least one modality remains available.
A higher \(r\) corresponds to a larger probability that each modality is unavailable.
All methods are evaluated with the same generated masks under each seed, so performance differences are not caused by different missing-pattern samples.
Results are reported separately for each missing rate to show how performance changes as modality incompleteness increases.

\subsection{Training Views}
\label{app:training_views}
During student training, we construct two types of incomplete views from each clean training sample. Fixed-missing views are generated by sampling one predefined non-empty available-modality subset, while random-missing views are generated by separately sampling a missing ratio from \(0.1\) to \(0.7\) and then drawing sample-wise modality-availability masks. All-missing random masks are rejected to ensure that at least one modality remains available. The student is optimized on these incomplete views, whereas the frozen complete-view teacher always receives the clean language, acoustic, and visual observations. The teacher provides prediction-level, fused-state, and modality-specific-state references, and no additional clean-view student-only branch is used.

\subsection{Nonverbal Conflict View Construction}
\label{app:conflict_construction}

In NCE, we construct conflict views at training time by replacing nonverbal modalities while keeping the target language unchanged. For a target sample \(i\), we sample a donor sample \(j\) from the same mini-batch with \(y_j\neq y_i\). We then sample a replaced modality set \(B\in\{\{\mathrm a\},\{\mathrm v\},\{\mathrm a,\mathrm v\}\}\). The conflict view keeps the target language feature and replaces the modalities in \(B\) with the corresponding donor features:
\[
\widetilde{\mathbf x}_{m,i}^{B}
=
\begin{cases}
\mathbf x_{m,j}, & m\in B,\\
\mathbf x_{m,i}, & m\notin B.
\end{cases}
\]
The supervision label remains the target label \(y_i\). The donor label \(y_j\) is used only to ensure label mismatch during construction and to compute conflict-analysis diagnostics such as donor attraction or rejection rate. If no different-label donor is available in the mini-batch, the target sample is skipped for NCE.

This construction is used as a training-time stress regularizer for target-preserving misleading nonverbal evidence. It provides a controlled perturbation setting rather than an exhaustive simulation of all real conversational asynchrony patterns.

\subsection{Optimization, Seeds, and Hardware}
\label{app:training_details}

\paragraph{Optimization.}
All main experiments use the Adam optimizer with learning rate \(1\mathrm{e}{-5}\), weight decay \(1\mathrm{e}{-5}\), and batch size \(16\).
IEMOCAP and MELD models are trained for \(150\) epochs.
CMU-MOSEI models are trained for \(100\) epochs in the supplementary setting.

\paragraph{Loss weights.}
The KD temperature is set to \(\tau=2\).
The loss weights are
\(\lambda_{\mathrm{kd}}=1.0\),
\(\lambda_{\mathrm{state}}=0.5\),
\(\lambda_{\mathrm{mstate}}=0.5\), and
\(\lambda_{\mathrm{NCE}}=1.0\).
NCE views are sampled with probability \(0.2\).
In the main implementation, NCE uses only the target-label conflict-view CE term; conflict-view KD and explicit margin losses are not used.

\paragraph{Training schedule and seeds.}
We train all models for the full schedule and apply the same predefined checkpoint-selection rule across compared methods.
Unless otherwise specified, all IEMOCAP, MELD, and CMU-MOSEI results are reported as the mean over five random seeds.
CMU-MOSEI is included as a supplementary utterance-level generalization check.

\paragraph{Hardware.}
All experiments are conducted on four NVIDIA RTX A5000 24GB GPUs.

\subsection{Inference}
\label{app:inference}

At inference time, only the student model is used.
The complete-view teacher, unavailable-state decoders, and NCE donor construction are used only during training and are discarded at inference.
Thus, CoRe-KD remains a single-student inference model under all missing-modality conditions.

\section{Baseline Fairness Control}
\label{app:baseline}

\begin{table*}[t]
\centering
\scriptsize
\setlength{\tabcolsep}{2.8pt}
\renewcommand{\arraystretch}{1.05}
\resizebox{0.98\textwidth}{!}{%
\begin{tabular}{@{}l|cc|cc|cc|cc@{}}
\toprule
\multirow{2}{*}{Method}
& \multicolumn{2}{c|}{\textbf{IEMOCAP-4}}
& \multicolumn{2}{c|}{\textbf{IEMOCAP-6}}
& \multicolumn{2}{c|}{\textbf{MELD-7}}
& \multicolumn{2}{c}{\textbf{CMU-MOSEI}} \\
\cmidrule(lr){2-3}
\cmidrule(lr){4-5}
\cmidrule(lr){6-7}
\cmidrule(lr){8-9}
& Fixed Avg. & RMFM Avg.
& Fixed Avg. & RMFM Avg.
& Fixed Avg. & RMFM Avg.
& Fixed Avg. & RMFM Avg. \\
\midrule

IMDer
& 68.35 $\pm$ 1.67 & 76.14 $\pm$ 1.02
& 56.19 $\pm$ 1.81 & 61.55 $\pm$ 0.68
& 46.10 $\pm$ 0.22 & 54.71 $\pm$ 0.20
& 76.83 $\pm$ 1.64 & 80.78 $\pm$ 2.38 \\

LNLN
& 70.02 $\pm$ 1.11 & 77.29 $\pm$ 0.90
& 57.32 $\pm$ 0.17 & 62.79 $\pm$ 0.85
& 52.66 $\pm$ 0.36 & 55.76 $\pm$ 0.25
& 77.80 $\pm$ 0.06 & 79.70 $\pm$ 0.23 \\

Corr-KD
& 74.07 $\pm$ 1.72 & 83.98 $\pm$ 0.85
& 58.40 $\pm$ 2.60 & 64.03 $\pm$ 1.30
& 51.35 $\pm$ 0.16 & 55.54 $\pm$ 0.08
& 73.19 $\pm$ 0.16 & 74.57 $\pm$ 0.73 \\

MoMKE
& 71.85 $\pm$ 1.38 & 77.22 $\pm$ 2.05
& 57.18 $\pm$ 1.27 & 62.45 $\pm$ 1.93
& 51.02 $\pm$ 1.29 & 54.98 $\pm$ 0.38
& 77.60 $\pm$ 0.44 & 79.62 $\pm$ 0.32 \\

MCULoRA
& 71.46 $\pm$ 2.47 & 79.89 $\pm$ 0.30
& 59.28 $\pm$ 0.83 & 64.82 $\pm$ 0.79
& 51.85 $\pm$ 0.11 & 55.16 $\pm$ 0.12
& 77.05 $\pm$ 0.66 & 81.17 $\pm$ 0.07 \\

ComP
& 70.35 $\pm$ 1.20 & 79.86 $\pm$ 0.91
& 58.72 $\pm$ 1.38 & 64.69 $\pm$ 0.83
& 52.12 $\pm$ 0.28 & 55.33 $\pm$ 0.26
& 77.81 $\pm$ 0.34 & 82.46 $\pm$ 0.47 \\

\textbf{CoRe-KD}
& \textbf{80.85 $\pm$ 1.83} & \textbf{86.48 $\pm$ 0.71}
& \textbf{62.85 $\pm$ 1.41} & \textbf{71.04 $\pm$ 0.76}
& \textbf{54.56 $\pm$ 1.05} & \textbf{57.84 $\pm$ 1.35}
& \textbf{79.97 $\pm$ 0.63} & \textbf{82.61 $\pm$ 0.23} \\

\bottomrule
\end{tabular}%
}
\caption{
Seed-level stability under the main and supplementary missing-modality evaluations.
Fixed Avg. is averaged over fixed-missing modality subsets, and RMFM Avg. is averaged over random-missing rates.
Scores are reported as mean $\pm$ standard deviation over five random seeds.
For IEMOCAP and MELD, the score is weighted F1; for CMU-MOSEI, we report the corresponding non-zero sentiment F1.
}
\label{tab:seed_variance_all}
\end{table*}

\subsection{Reproduction Setting}

We compare CoRe-KD with representative incomplete-modality and robust MER baselines under the same fixed-missing and random-missing evaluation protocols.
For reproduced methods, we retrain all models with the same data splits, checkpoint selection rule, and evaluation settings, and use the same feature inputs whenever feature replacement is supported.
Whenever possible, we follow the original papers and released implementations; otherwise, we reimplement the method according to the original description and tune hyperparameters on the development split when available; for settings without a development split, we use fixed hyperparameters under the same protocol.
Using shared extracted language, acoustic, and visual features whenever supported controls for feature-specific effects and makes the comparison focus on the multimodal learning strategy.

\subsection{Baseline Categories}

The compared baselines in Table~\ref{tab:missing_results_iemocap6_meld7} cover three representative families according to their primary modeling emphasis.
IMDer and LNLN represent reconstruction- or recovery-oriented approaches.
Corr-KD and MoMKE represent distillation- or expert-knowledge-transfer-based approaches.
MCULoRA and ComP represent adaptation- and prompting-based approaches for incomplete multimodal learning.
Together, these baselines cover recent major lines of missing-modality multimodal emotion recognition and provide a broad comparison for CoRe-KD.

\subsection{Feature and Teacher Control}

\paragraph{Baseline and feature control.}
To ensure controlled comparisons, reproduced baselines use the same language, acoustic, and visual features as CoRe-KD whenever feature replacement is supported.
For methods whose official implementations are tightly coupled with their original feature preprocessing, we keep their original-feature setting to avoid feature-incompatibility artifacts while applying the same checkpoint selection rule.
This reduces the risk of underestimating baselines due to feature incompatibility while keeping the main comparison as controlled as possible.
Thus, performance differences are less likely to come from feature-specific advantages and more directly reflect the multimodal learning strategy.

\paragraph{Controlled distillation comparison.}
For distillation-based comparisons, we use the same clean complete-view teacher whenever applicable.
This includes the vanilla prediction-level KD, DKD, and Corr-KD variants in the teacher-backbone generalization study.
All distillation variants are trained with the same incomplete-view data and evaluated under the same fixed- and random-missing protocols as CoRe-KD.
These comparisons test whether structured complete-view state anchoring provides useful supervision beyond prediction-level transfer in our robust MER setting.
By sharing the same teacher source, the comparison isolates the effect of CoRe-KD's structured anchoring objective from the effect of teacher quality.

\subsection{Training Schedule and Seeds}

All compared methods are trained under the same main experimental protocol whenever applicable.
We train each method for the full schedule with its recommended or tuned hyperparameters, and apply the same predefined checkpoint selection rule across fixed-missing and random-missing evaluations.
Unless otherwise specified, results on IEMOCAP, MELD, and CMU-MOSEI are averaged over five random seeds.
For each seed, all methods are evaluated using the same fixed-missing conditions and the same generated random-missing masks under the corresponding protocols.
This unified training and evaluation protocol reduces variance from implementation choices and makes the reported differences more comparable across methods.

\subsection{Stability Across Random Seeds}
\label{app:seed_variance}

\paragraph{Seed-level stability.}
Table~\ref{tab:seed_variance_all} reports mean and standard deviation over five random seeds for the main fixed-missing and random-missing averages.
Fixed Avg. is averaged over all fixed available-modality subsets, and RMFM Avg. is averaged over all evaluated random-missing rates.
All methods use the same fixed-missing conditions and, for random-missing evaluations, the same generated masks under each seed, so the variance mainly reflects training stochasticity and random initialization.

\paragraph{Analysis.}
CoRe-KD achieves the highest average F1 across IEMOCAP-4, IEMOCAP-6, MELD-7, and CMU-MOSEI under both protocols.
On the dialogue-context benchmarks, it improves over the strongest baseline by \(6.78/2.50\), \(3.57/6.22\), and \(1.90/2.08\) points on IEMOCAP-4, IEMOCAP-6, and MELD-7 under Fixed/RMFM Avg., respectively.
On CMU-MOSEI, CoRe-KD obtains the best fixed-missing average and remains comparable under random missing.
Overall, the gains are larger than typical seed-level fluctuations, with standard deviations comparable to reproduced baselines, suggesting that the improvements are stable across seeds.
\section{Teacher Architecture}
\label{app:architecture}

\subsection{Overview}

The complete-view teacher architecture and Gaussian-inspired state-head implementation are described below.
For IEMOCAP and MELD, the hidden dimension is set to \(1024\); for CMU-MOSEI, it is set to \(256\).
The Gaussian-inspired state dimension is \(d=256\).
Each intra-/inter-modal Transformer uses one layer and \(8\) attention heads.
The dropout rate is \(0.5\) for IEMOCAP and MELD, and \(0.3\) for CMU-MOSEI.

We do not use a sliding dialogue window.
Instead, each dialogue-context sequence or utterance-level input sequence is padded and processed as a whole sequence.
The maximum positional length is \(512\), and the maximum observed sequence lengths are \(110\), \(33\), and \(87\) for IEMOCAP, MELD, and CMU-MOSEI, respectively.

\subsection{Complete-View Teacher}
\label{app:teacher_model}

CoRe-KD uses a frozen complete-view teacher \(T\) to provide structured references for student training.
The teacher is trained only with clean complete language, acoustic, and visual observations, and is fixed after convergence.
In our implementation, \(T\) is instantiated as a Transformer-based multimodal dialogue encoder, following the intra-/inter-modal Transformer encoding design in prior multimodal sentiment and conversational emotion recognition models~\citep{yuan2021transformer,ma2024sdt}.
The teacher is used only as a complete-view reference model; CoRe-KD does not rely on teacher-specific reconstruction or self-distillation objectives.

For a dialogue context \(\mathcal D_i\) around target utterance \(u_i\), let
\[
\mathbf X_{m,i}^{t}
=
\{\mathbf x_{k}^{m,t}\mid k\in\mathcal C_i\},
\qquad
m\in\mathcal M=\{\mathrm \ell,\mathrm a,\mathrm v\},
\]
where \(\mathcal C_i\) denotes the utterance-index set of the input sequence containing \(u_i\).
Each modality sequence is projected into a shared hidden space and augmented with positional and speaker/context embeddings:
\begin{equation}
\mathbf Z_{m,i}^{t}
=
\operatorname{Proj}_{m}^{t}
\left(
\mathbf X_{m,i}^{t}
\right)
+
\mathbf P_i
+
\mathbf S_i .
\end{equation}
Here, \(\mathbf P_i\) and \(\mathbf S_i\) denote positional and speaker/context embeddings; \(\mathbf S_i\) is omitted for utterance-level datasets without speaker information.

The teacher first applies an intra-modal Transformer to encode contextual dynamics within each modality:
\begin{equation}
\mathbf U_{m,i}^{t}
=
\operatorname{IntraTrans}_{m}^{t}
\left(
\mathbf Z_{m,i}^{t}
\right),
\qquad
m\in\mathcal M .
\end{equation}
Here, \(\operatorname{IntraTrans}_{m}^{t}(\cdot)\) is a self-attention Transformer encoder.

Following the common inter-modal Transformer design, modality \(m\) further attends to modality \(n\).
Let \(\mathcal I_{mn}^{t}\) denote this inter-modal Transformer, where the first input provides queries and the second input provides keys and values:
\begin{equation}
\mathbf V_{mn,i}^{t}
=
\mathcal I_{mn}^{t}
\left(
\mathbf U_{m,i}^{t},
\mathbf U_{n,i}^{t}
\right),
\qquad
n\neq m .
\end{equation}
This produces cross-modal enhanced sequences \(\{\mathbf V_{mn,i}^{t}\}_{n\neq m}\) for each target modality \(m\).

For target utterance \(u_i\), we select the target-position representations from the intra-modal and inter-modal sequences.
Let \(\operatorname{Sel}_i(\cdot)\) denote target-position selection:
\begin{equation}
\begin{aligned}
\mathbf r_{m,i}^{t}
&=
\operatorname{Sel}_i
\left(
\mathbf U_{m,i}^{t}
\right),\\
\mathbf r_{mn,i}^{t}
&=
\operatorname{Sel}_i
\left(
\mathbf V_{mn,i}^{t}
\right).
\end{aligned}
\end{equation}

Inspired by hierarchical gated fusion, we combine the self-contextual and cross-modal representations by a lightweight gate.
Let \(\mathbf c_{m,i}^{t}\) be the concatenation of \(\mathbf r_{m,i}^{t}\) and \(\{\mathbf r_{mn,i}^{t}\}_{n\neq m}\).
Then,
\begin{equation}
\begin{aligned}
\mathbf g_{m,i}^{t}
&=
\operatorname{softmax}
\left(
\mathbf W_g^t \mathbf c_{m,i}^{t}
+
\mathbf b_g^t
\right),\\
\mathbf h_{m,i}^{t}
&=
g_{m,0}^{t}\mathbf r_{m,i}^{t}
+
\sum_{n\neq m}
g_{m,n}^{t}\mathbf r_{mn,i}^{t}.
\end{aligned}
\end{equation}
Here, \(\mathbf h_{m,i}^{t}\) is the Transformer-enhanced teacher-side modality representation.

\subsection{Gaussian-inspired State Head}

Gaussian-inspired state heads are attached after these Transformer-enhanced modality representations.
In the main text, \(\operatorname{StateHead}_{m}\) denotes the abstract state-head interface.
In this teacher implementation, it is instantiated by a lightweight Gaussian head \(\operatorname{GHead}_{m}^{t}\).
For each modality, \(\operatorname{GHead}_{m}^{t}\) outputs a \(2d\)-dimensional vector, which is split into a state location and a log-variance vector:
\begin{equation}
\left(
\boldsymbol\mu_{m,i}^{t},
\boldsymbol\ell_{m,i}^{t}
\right)
=
\operatorname{Split}
\left(
\operatorname{GHead}_{m}^{t}
\left(
\mathbf h_{m,i}^{t}
\right)
\right),
\end{equation}
where
\(\boldsymbol\mu_{m,i}^{t},\boldsymbol\ell_{m,i}^{t}\in\mathbb R^d\).
The Gaussian head is implemented as a lightweight MLP with normalization, nonlinearity, dropout, and a final linear projection.

For numerical stability, we clamp the log-variance to
\([\ell_{\min},\ell_{\max}]\), where \(\ell_{\min}=-6\) and \(\ell_{\max}=2\):
\begin{equation}
\boldsymbol\ell_{m,i}^{t}
=
\operatorname{clamp}_{[\ell_{\min},\ell_{\max}]}
\left(
\boldsymbol\ell_{m,i}^{t}
\right).
\end{equation}
The diagonal scale and precision are computed as
\begin{equation}
\begin{aligned}
\boldsymbol\sigma_{m,i}^{t}
&=
\exp
\left(
\tfrac{1}{2}
\boldsymbol\ell_{m,i}^{t}
\right),\\
\boldsymbol\kappa_{m,i}^{t}
&=
\exp
\left(
-\boldsymbol\ell_{m,i}^{t}
\right).
\end{aligned}
\end{equation}
The modality-specific teacher state is
\begin{equation}
q_{m,i}^{t}
=
\mathcal G
\left(
\boldsymbol\mu_{m,i}^{t},
\boldsymbol\sigma_{m,i}^{t}
\right).
\end{equation}

\subsection{Product-of-Experts Fusion}

The complete-view fused teacher state is obtained by Product-of-Experts fusion over all modality states:
\begin{equation}
\begin{aligned}
(\boldsymbol\sigma_i^t)^2
&=
\left(
\mathbf 1
+
\sum_{m\in\mathcal M}
\boldsymbol\kappa_{m,i}^{t}
\right)^{-1},\\
\boldsymbol\mu_i^t
&=
(\boldsymbol\sigma_i^t)^2
\odot
\sum_{m\in\mathcal M}
\boldsymbol\kappa_{m,i}^{t}
\odot
\boldsymbol\mu_{m,i}^{t},\\
q_i^t
&=
\mathcal G
\left(
\boldsymbol\mu_i^t,
\boldsymbol\sigma_i^t
\right).
\end{aligned}
\end{equation}
The \(\mathbf 1\) term corresponds to a unit Gaussian prior precision.
It provides a weak fixed reference precision and is kept identical across teacher and student variants.

The teacher predicts emotions from the fused state location:
\begin{equation}
\mathbf o_i^t
=
C^t
\left(
\boldsymbol\mu_i^t
\right),
\qquad
p_i^t
=
\operatorname{softmax}
\left(
\mathbf o_i^t
\right).
\end{equation}
It is trained with standard complete-view classification:
\begin{equation}
\mathcal L_T
=
\mathbb E_i
\left[
\operatorname{CE}
\left(
y_i,
\mathbf o_i^t
\right)
\right].
\end{equation}

After training, all teacher parameters are frozen.
During student training, the teacher always receives the clean complete view and provides the deterministic structured reference:
\begin{equation}
\mathcal T_i
=
\left(
\mathbf o_i^t,
p_i^t,
q_i^t,
\{q_{m,i}^{t}\}_{m\in\mathcal M}
\right).
\end{equation}
The student uses \(\mathcal T_i\) for prediction anchoring, fused-state anchoring, and unavailable-modality-state anchoring.

\subsection{Student State Heads and Unavailable-State Decoders}

The teacher state heads, student state heads, and unavailable-state decoders use the same log-variance Gaussian-inspired state parameterization, but all parameters are independent.
For an unavailable-state decoder \(D_m\), the output is split into a mean vector and a log-variance vector, followed by the same clipping and scale computation.
The unavailable-state decoders are used only during student training and are removed at inference time.
\section{Theoretical Notes}
\label{app:theory}

This section provides lightweight theoretical notes for the main design choices of CoRe-KD.
The goal is not to claim calibrated probabilistic inference or a formal generalization guarantee.
Instead, we use simplified analyses to clarify how complete-view state anchoring, location-scale matching, and NCE can provide structured supervision beyond prediction-level KD or raw reconstruction.

\subsection{PoE Fusion as Precision-Weighted State Estimation}
\label{app:theory_poe}

The Product-of-Experts fusion used in Eq.~(3) can be interpreted as the solution of a precision-weighted state estimation problem.
For an available modality set \(\mathcal A_i^c\), suppose each modality provides a location vector \(\boldsymbol\mu_{m,i}^{c}\) and a diagonal precision vector \(\boldsymbol\kappa_{m,i}^{c}\).
For compactness, define
\begin{equation}
\mathbf r_{m,i}^c(\boldsymbol z)
=
\sqrt{\boldsymbol\kappa_{m,i}^{c}}
\odot
(\boldsymbol z-\boldsymbol\mu_{m,i}^{c}),
\end{equation}
where the square root and product are element-wise.
Consider the objective
\begin{equation}
\mathcal J(\boldsymbol z)
=
\frac{1}{2}\|\boldsymbol z\|_2^2
+
\frac{1}{2}
\sum_{m\in \mathcal A_i^c}
\|\mathbf r_{m,i}^c(\boldsymbol z)\|_2^2 .
\end{equation}
The first term corresponds to a unit Gaussian prior.
Setting the gradient to zero gives
\begin{equation}
\boldsymbol\Lambda_i^c
\odot
\boldsymbol z_i^\star
=
\boldsymbol\eta_i^c,
\end{equation}
where
\begin{equation}
\begin{aligned}
\boldsymbol\Lambda_i^c
&=
\mathbf 1+
\sum_{m\in \mathcal A_i^c}
\boldsymbol\kappa_{m,i}^{c},\\
\boldsymbol\eta_i^c
&=
\sum_{m\in \mathcal A_i^c}
\boldsymbol\kappa_{m,i}^{c}
\odot
\boldsymbol\mu_{m,i}^{c}.
\end{aligned}
\end{equation}
Therefore,
\begin{equation}
\begin{aligned}
\boldsymbol z_i^\star
&=
(\boldsymbol\Lambda_i^c)^{-1}
\odot
\boldsymbol\eta_i^c,\\
(\boldsymbol\sigma_i^c)^2
&=
(\boldsymbol\Lambda_i^c)^{-1}.
\end{aligned}
\end{equation}
This is exactly the fused state location and variance in the PoE fusion rule.
This interpretation clarifies that unavailable modalities are omitted from the summation, while modalities assigned higher precision have larger influence on the fused location.
The unit prior term acts as a weak reference precision and helps stabilize fusion when only a few modalities are available.

\subsection{Location-Scale Matching Geometry}
\label{app:theory_state_distance}

CoRe-KD does not require the Gaussian-inspired states to be calibrated posteriors.
We use them as a structured location-scale interface for teacher-reference matching.
For two diagonal Gaussian-inspired states
\[
q=\mathcal G(\boldsymbol\mu,\boldsymbol\sigma),
\qquad
q'=\mathcal G(\boldsymbol\mu',\boldsymbol\sigma'),
\]
the squared 2-Wasserstein distance has the closed form
\begin{equation}
W_2^2(q,q')
=
\|\boldsymbol\mu-\boldsymbol\mu'\|_2^2
+
\|\boldsymbol\sigma-\boldsymbol\sigma'\|_2^2 .
\end{equation}
Thus, the state distance used in CoRe-KD can be written as
\begin{equation}
\mathcal D_G(q,q')
=
\frac{1}{d}
W_2^2(q,q').
\end{equation}
Compared with matching only \(\boldsymbol\mu\), this distance also encourages the student to match the teacher-side scale structure.

This state distance is also connected to student-side prediction consistency under a fixed classifier.
Let
\[
\boldsymbol\mu_s=\boldsymbol\mu_i^{s,c},
\qquad
\boldsymbol\mu_t=\boldsymbol\mu_i^{t}.
\]
If the student classifier \(C^s(\cdot)\) is \(L\)-Lipschitz with respect to the fused state location, then
\begin{equation}
\begin{aligned}
\|C^s(\boldsymbol\mu_s)-C^s(\boldsymbol\mu_t)\|_2
&\le
L\|\boldsymbol\mu_s-\boldsymbol\mu_t\|_2\\
&\le
L\sqrt{d\,\mathcal D_G(q_i^{s,c},q_i^t)} .
\end{aligned}
\end{equation}
Therefore, reducing the fused-state distance constrains the logit change induced by the student's fused-state location.
Prediction-level anchoring separately aligns the student output with the complete-view teacher logits.
This does not by itself guarantee higher classification accuracy, but it provides one explanation for why state anchoring can connect complete-view teacher references with incomplete-view predictions.

\subsection{Why Raw Feature Reconstruction Can Be Ambiguous}
\label{app:theory_reconstruction}

A common way to handle missing modalities is to reconstruct missing raw or feature-level inputs.
However, under squared reconstruction loss, the optimal predictor is the conditional mean of the missing modality given the observed modalities.
Let \(\mathbf X_{\mathrm{obs}}\) denote observed modalities and \(\mathbf X_{\mathrm{mis}}\) denote missing modalities.
The reconstruction objective is
\begin{equation}
\min_g
\mathbb E
\left[
\|g(\mathbf X_{\mathrm{obs}})
-
\mathbf X_{\mathrm{mis}}\|_2^2
\right].
\end{equation}
Its Bayes optimal solution is
\begin{equation}
g^\star(\mathbf X_{\mathrm{obs}})
=
\mathbb E[
\mathbf X_{\mathrm{mis}}
\mid
\mathbf X_{\mathrm{obs}}
].
\end{equation}
When the conditional distribution
\(p(\mathbf X_{\mathrm{mis}}\mid \mathbf X_{\mathrm{obs}})\)
is multi-modal, this conditional mean may average over several plausible completions.
Such an averaged reconstruction may not correspond to a realistic modality observation and may dilute label-relevant nonverbal evidence.

This analysis does not imply that reconstruction is always ineffective.
Rather, it motivates using complete-view teacher states as task-oriented targets when raw missing inputs are not uniquely identifiable.
CSA avoids requiring a unique raw completion and instead asks the student to match complete-view fused and modality-specific state references derived from the teacher's task prediction pathway.

\subsection{NCE as Target-over-Donor Margin Regularization}
\label{app:theory_nce}

NCE constructs target-preserving conflict views by keeping the target language unchanged while replacing nonverbal modalities with a different-label donor.
Let \(y_i\) be the target label and \(y_j\neq y_i\) be the donor label.
For a conflict view with replaced modality set \(b\), let
\(\widetilde{\mathbf o}_{i}^{s,b}\)
be the student logits.
Define the target-over-donor margin as
\begin{equation}
\Delta_{ij}^{b}
=
\widetilde{o}_{i,y_i}^{s,b}
-
\widetilde{o}_{i,y_j}^{s,b}.
\end{equation}
For compactness, define
\begin{equation}
\delta_{ik}^{b}
=
\widetilde{o}_{i,k}^{s,b}
-
\widetilde{o}_{i,y_i}^{s,b}.
\end{equation}
The conflict-view cross-entropy is
\begin{equation}
\operatorname{CE}_{i}^{b}
=
\log
\left(
1+
\sum_{k\neq y_i}
\exp(\delta_{ik}^{b})
\right).
\end{equation}
Since \(y_j\neq y_i\), the donor-label term is included in the summation.
Therefore,
\begin{equation}
\operatorname{CE}_{i}^{b}
\ge
\log
\left(
1+\exp(-\Delta_{ij}^{b})
\right).
\end{equation}
Minimizing the conflict-view CE therefore implicitly encourages a larger target-over-donor margin.
This gives a margin-based interpretation of NCE: the model is trained to keep the target label preferred even when nonverbal cues are replaced by donor-label evidence.

This result only explains the optimization effect of NCE under the constructed conflict views.
It does not claim that donor swapping exhaustively models all real conversational asynchrony or nonverbal mismatch.
In our experiments, NCE is used as a controlled stress regularizer for misleading nonverbal evidence under the evaluated construction.
\section{Controlled State-Interface Comparison}
\label{app:controlled_state}

This section provides controlled comparisons with simpler auxiliary and teacher-guided objectives.
The goal is to examine whether the gains of CoRe-KD come from structured complete-view state anchoring rather than from prediction-level KD, generic hidden-state regression, raw feature reconstruction, or regularizer stacking.
All variants in this section use the same backbone, input features, missing-modality training views, checkpoint selection rule, and evaluation masks.
Unless otherwise specified, the comparison is conducted on IEMOCAP-6, which is the main controlled-analysis dataset.

\subsection{Compared Variants}

\paragraph{KD only.}
The student is trained with task CE and prediction-level KD from the frozen complete-view teacher.

\paragraph{KD + Hidden MSE.}
The student additionally regresses the teacher hidden representation using an MSE loss.
This variant tests whether generic representation matching can explain the gains.

\paragraph{KD + Raw Feature Reconstruction.}
The student additionally reconstructs missing modality features from the available-view representation.
This variant tests whether raw feature reconstruction is a sufficient target under the same backbone and training recipe.

\paragraph{KD + Deterministic State.}
The student matches deterministic state locations without modeling the scale term.
This variant tests whether point-vector state regression is sufficient for complete-view guidance.

\paragraph{KD + Gaussian CSA.}
The student matches both the teacher state location and scale through the Gaussian-inspired state distance.
This variant isolates complete-view state anchoring without unavailable-modality-state anchoring or NCE.

\paragraph{CoRe-KD.}
The full model combines prediction anchoring, Gaussian complete-view state anchoring, unavailable-modality-state anchoring, and NCE.

\begin{table*}[t]
\centering
\small
\setlength{\tabcolsep}{2.6pt}
\renewcommand{\arraystretch}{1.05}
\resizebox{0.99\textwidth}{!}{%
\begin{tabular}{@{}l|ccccccc|ccc@{}}
\toprule
Variant
& KD & Hidden MSE & Raw Rec. & Det. State & Gaussian CSA & M-State & NCE
& Fixed Avg F1 & RMFM Avg F1 & High-Missing Avg F1 \\
\midrule
KD only
& Yes & No & No & No & No & No & No
& 56.89 $\pm$ 1.26 & 68.12 $\pm$ 0.46 & 65.52 $\pm$ 0.52 \\

KD + Hidden MSE
& Yes & Yes & No & No & No & No & No
& 58.62 $\pm$ 0.93 & 68.41 $\pm$ 0.53 & 65.84 $\pm$ 0.86 \\

KD + Raw Feature Rec.
& Yes & No & Yes & No & No & No & No
& 59.24 $\pm$ 0.59 & 69.30 $\pm$ 0.58 & 66.89 $\pm$ 0.75 \\

KD + Deterministic State
& Yes & No & No & Yes & No & No & No
& 59.63 $\pm$ 0.25 & 69.43 $\pm$ 0.40 & 67.14 $\pm$ 0.99 \\

KD + Gaussian CSA
& Yes & No & No & No & Yes & No & No
& 59.83 $\pm$ 0.18 & 69.68 $\pm$ 0.75 & 67.46 $\pm$ 1.36 \\

CoRe-KD
& Yes & No & No & No & Yes & Yes & Yes
& \textbf{62.85 $\pm$ 1.41} & \textbf{71.04 $\pm$ 0.76} & \textbf{69.26 $\pm$ 1.32} \\
\bottomrule
\end{tabular}%
}
\caption{
Controlled comparison with simpler auxiliary and teacher-guided objectives on IEMOCAP-6.
Fixed Avg denotes the average weighted F1 over fixed-missing conditions.
RMFM Avg denotes the average weighted F1 over random-missing rates.
High-Missing Avg denotes the average weighted F1 over high missing rates, i.e., \(r\in\{0.5,0.6,0.7\}\).
Scores are reported as mean \(\pm\) standard deviation over five random seeds.
M-State denotes unavailable-modality-state anchoring.
The full CoRe-KD row is aligned with the final main-setting aggregate.
}
\label{tab:controlled_state_interface}
\end{table*}

\subsection{Analysis}

Table~\ref{tab:controlled_state_interface} compares prediction-level KD with stronger teacher-guided objectives under the same IEMOCAP-6 controlled-analysis setting.
KD only is already a strong baseline, while hidden-state regression and raw feature reconstruction provide additional gains.
State-based variants further improve the three averaged metrics, with Gaussian CSA outperforming deterministic state matching by matching both location and scale.
Full CoRe-KD achieves the best results, improving over KD only by 5.96, 2.92, and 3.74 F1 points on Fixed Avg, RMFM Avg, and High-Missing Avg, respectively.
It also improves over Gaussian CSA by 3.02, 1.36, and 1.80 points.
These results support that the final gains come from structured complete-view state anchoring and complementary robustness regularization, rather than from prediction-level KD or generic auxiliary matching alone.

\subsection{Location-Only vs. Location-and-Scale Matching}
\label{app:mu_sigma_ablation}

To further examine the effect of the Gaussian-inspired parameterization, we compare matching only the state location \(\boldsymbol\mu\) against matching both location and scale \((\boldsymbol\mu,\boldsymbol\sigma)\).
Both variants use the same full CoRe-KD setting except for the state-matching distance.
The location-only variant replaces the Gaussian-inspired state distance with an MSE loss on \(\boldsymbol\mu\), while the location-and-scale variant uses the same Gaussian-inspired state distance as the final CoRe-KD model.

\begin{table}[t]
\centering
\scriptsize
\setlength{\tabcolsep}{2.8pt}
\renewcommand{\arraystretch}{1.08}
\resizebox{\linewidth}{!}{%
\begin{tabular}{@{}l|ccc@{}}
\toprule
Variant & Fixed Avg F1 & RMFM Avg F1 & High-Missing Avg F1 \\
\midrule
Match \(\boldsymbol\mu\) only
& 60.22 $\pm$ 0.43
& 69.65 $\pm$ 0.40
& 67.54 $\pm$ 0.25 \\
Match \(\boldsymbol\mu,\boldsymbol\sigma\)
& \textbf{62.85 $\pm$ 1.41}
& \textbf{71.04 $\pm$ 0.76}
& \textbf{69.26 $\pm$ 1.32} \\
\bottomrule
\end{tabular}%
}
\caption{
Effect of matching only the state location versus matching both state location and scale on IEMOCAP-6.
All columns report average F1: Fixed Avg over fixed-missing conditions, RMFM Avg over random-missing rates, and High-Missing Avg over \(r\in\{0.5,0.6,0.7\}\).
Scores are reported as mean \(\pm\) standard deviation over five random seeds.
The \((\boldsymbol\mu,\boldsymbol\sigma)\) row corresponds to the final CoRe-KD state-matching distance and is aligned with the main aggregate results.
}
\label{tab:mu_sigma_ablation}
\end{table}

Table~\ref{tab:mu_sigma_ablation} isolates the role of the scale term in the Gaussian-inspired state distance.
Matching both \(\boldsymbol\mu\) and \(\boldsymbol\sigma\) improves over location-only matching by 2.63, 1.39, and 1.72 F1 points on Fixed Avg, RMFM Avg, and High-Missing Avg, respectively.
This suggests that the scale term provides useful information beyond point-vector location matching under the same final CoRe-KD setting.
\section{Nonverbal Conflict Analysis}
\label{app:conflict_analysis}

\subsection{Controlled Conflict Evaluation}

We further evaluate CoRe-KD under controlled target-preserving nonverbal conflict views.
For each target utterance, the language modality is kept unchanged, while acoustic, visual, or acoustic-visual observations are replaced by different-label donor samples.
The target label is preserved during evaluation.
This protocol provides a controlled diagnostic for whether the student remains anchored to the target utterance when nonverbal observations introduce misleading donor-label evidence.
Unlike the state-space rejection metric in the main analysis, this evaluation reports task-level F1 under conflict views.

\begin{table}[t]
\centering
\small
\setlength{\tabcolsep}{3.5pt}
\renewcommand{\arraystretch}{1.05}
\begin{tabular}{@{}l|ccc|c@{}}
\toprule
Method & A-conflict & V-conflict & AV-conflict & Avg \\
\midrule
KD only
& 61.76 & 73.32 & 60.26 & 65.11 \\
w/o \(\mathcal L_{\mathrm{NCE}}\)
& 62.78 & 72.81 & 61.40 & 65.66 \\
CoRe-KD
& 63.69 & 72.87 & 62.71 & 66.42 \\
\bottomrule
\end{tabular}
\caption{
Controlled target-preserving nonverbal conflict results.
All columns report F1.
}
\label{tab:conflict_results}
\end{table}

As shown in Table~\ref{tab:conflict_results}, CoRe-KD achieves the best average F1 under the controlled conflict views.
Compared with KD only, CoRe-KD improves the average F1 from 65.11 to 66.42, with clear gains under A-conflict and AV-conflict.
Compared with the variant without \(\mathcal L_{\mathrm{NCE}}\), CoRe-KD further improves the average F1 from 65.66 to 66.42.
The V-conflict result is comparable to KD only rather than uniformly better, suggesting that the gain mainly comes from improved robustness to acoustic and audio-visual conflicts.
Overall, this task-level evaluation complements the state-space analysis and suggests that nonverbal conflict exposure can help the student preserve target-side evidence when misleading nonverbal cues are introduced.

\subsection{Donor Selection Sensitivity}

We additionally examine whether the conflict analysis depends on a single donor construction strategy.
Specifically, we compare random different-label donors with different-label donors further constrained to come from the same speaker, the same dialogue, or semantically close utterances.
This analysis is intended as a sensitivity check rather than as a claim that donor swapping exhaustively models real conversational asynchrony or all forms of nonverbal conflict.

\begin{table}[t]
\centering
\small
\setlength{\tabcolsep}{3.2pt}
\renewcommand{\arraystretch}{1.05}
\resizebox{\linewidth}{!}{%
\begin{tabular}{@{}l|ccc|c@{}}
\toprule
Donor Type & A-conflict & V-conflict & AV-conflict & Rej. Rate \\
\midrule
Random different-label donor
& 63.69 & 72.87 & 62.71 & 95.36 \\
Same-speaker donor
& 64.17 & 72.95 & 62.56 & 94.38 \\
Same-dialogue donor
& 63.46 & 73.34 & 62.87 & 82.61 \\
Semantically close donor
& 63.60 & 72.87 & 62.38 & 85.85 \\
\bottomrule
\end{tabular}%
}
\caption{
Sensitivity analysis for different donor selection strategies.
Conflict columns report F1, and Rej. Rate denotes the fraction of conflict-view states closer to the target anchor than to the donor anchor.
}
\label{tab:donor_sensitivity}
\end{table}

Table~\ref{tab:donor_sensitivity} shows that the task-level F1 remains stable across different donor construction strategies.
Averaged over the three conflict views, the F1 varies only slightly across random different-label, same-speaker, same-dialogue, and semantically close donors.
The rejection rate decreases for same-dialogue and semantically close donors, suggesting that these donors form harder conflict cases, but the corresponding F1 remains close to the random-donor setting.
These results suggest that the observed conflict robustness is not merely an artifact of one arbitrary donor choice, while leaving broader real-world nonverbal asynchrony as a more general setting beyond this controlled diagnostic.

\section{Mechanism Metrics}
\label{app:mechanism}

The metrics in this section are diagnostic measures for examining whether the learned states behave consistently with the intended anchoring behavior.
They are not used as independent proof of the method, since they are defined with respect to the same teacher-state interface used during training.
We therefore interpret these diagnostics together with the external F1 results and the controlled comparisons in Appendix~\ref{app:controlled_state}.

\subsection{Fused-State Drift}

We use fused-state drift to measure how far the student fused state deviates from the complete-view teacher state under incomplete observations.
For a test condition \(c\), fused-state drift is computed as
\begin{equation}
\mathrm{Drift}(c)
=
\mathbb E_i
\left[
\mathcal D_G
\left(
q_i^{s,c},
q_i^t
\right)
\right],
\end{equation}
where \(q_i^{s,c}\) is the student fused Gaussian-inspired state under condition \(c\), \(q_i^t\) is the complete-view teacher state, and \(\mathcal D_G(\cdot,\cdot)\) is the state distance defined in the main text.
Lower drift indicates closer alignment with the complete-view teacher reference.

\subsection{Conflict-View Anchors}

For target-preserving nonverbal conflict analysis, we use the complete-view teacher states of the target and donor samples as anchors.
Given a target sample \(i\), a different-label donor \(j\), and a replaced nonverbal modality set \(b\), the student produces a conflict-view state \(\widetilde q_i^{s,b}\).
Following the main analysis, we use the state-location distance
\(d_\mu(q,q')=\|\boldsymbol\mu(q)-\boldsymbol\mu(q')\|_2\)
and define the target-anchor and donor-anchor distances as
\begin{equation}
\begin{aligned}
d_{\mathrm{tar}}^{i,b}
&=
d_\mu
\left(
\widetilde q_i^{s,b},
q_i^t
\right),\\
d_{\mathrm{don}}^{i,j,b}
&=
d_\mu
\left(
\widetilde q_i^{s,b},
q_j^t
\right).
\end{aligned}
\end{equation}

\subsection{Rejection Rate}

The rejection indicator measures whether the conflict-view state remains closer to the target anchor than to the donor anchor:
\begin{equation}
r_{i,j}^{b}
=
\mathbb I
\left[
d_{\mathrm{tar}}^{i,b}
<
d_{\mathrm{don}}^{i,j,b}
\right].
\end{equation}
The rejection rate is the average of \(r_{i,j}^{b}\) over valid target--donor pairs and conflict types:
\begin{equation}
\mathrm{Rej}
=
\mathbb E_{i,j,b}
\left[
r_{i,j}^{b}
\right].
\end{equation}
A higher rejection rate indicates that the conflict-view state is more often closer to the target teacher anchor than to the donor teacher anchor.

\subsection{State and Logit Margins}

We also report the target-over-donor state-location margin:
\begin{equation}
m_{i,j}^{b}
=
d_{\mathrm{don}}^{i,j,b}
-
d_{\mathrm{tar}}^{i,b}.
\end{equation}
A positive margin means that the conflict-view state is closer to the target teacher anchor than to the donor teacher anchor.

For prediction-level conflict analysis, we use the target-versus-donor logit margin:
\begin{equation}
\Delta_{ij}^{b}
=
\widetilde o_{i,y_i}^{s,b}
-
\widetilde o_{i,y_j}^{s,b}.
\end{equation}
A larger \(\Delta_{ij}^{b}\) indicates lower donor-label attraction in the conflict view.

\section{Supplementary IEMOCAP-4 Results}
\label{app:iemocap4_supp}

\begin{table*}[t]
\centering
\small
\setlength{\tabcolsep}{2.4pt}
\renewcommand{\arraystretch}{0.98}
\setlength{\arrayrulewidth}{0.4pt}
\setlength{\doublerulesep}{1.2pt}

\providecommand{\ourscell}{\cellcolor{gray!12}}
\providecommand{\best}[1]{\textbf{#1}}
\providecommand{\second}[1]{\underline{#1}}

\resizebox{0.99\textwidth}{!}{%
\begin{tabular}{@{}c|l||*{7}{c}@{}}
\Xhline{0.9pt}
\multicolumn{2}{c||}{Testing Condition}
& $\{\mathrm a\}$ & $\{\mathrm v\}$ & $\{\mathrm \ell\}$
& $\{\mathrm \ell,\mathrm a\}$ & $\{\mathrm a,\mathrm v\}$
& $\{\mathrm \ell,\mathrm v\}$ & $\{\mathrm \ell,\mathrm a,\mathrm v\}$ \\
\hline
Dataset & Method
& Acc(\%)/F1(\%) & Acc(\%)/F1(\%) & Acc(\%)/F1(\%)
& Acc(\%)/F1(\%) & Acc(\%)/F1(\%) & Acc(\%)/F1(\%)
& Acc(\%)/F1(\%) \\
\hline

\multirow{8}{*}{\begin{tabular}{c}IEMOCAP\\Four\end{tabular}}
& IMDer
& 57.07/55.31 & 44.12/38.28 & 78.97/78.87
& 82.03/82.10 & 61.40/60.64 & 80.54/80.39 & 82.85/82.86 \\
& Corr-KD
& \second{68.61}/\second{68.20} & 43.98/35.29 & \second{84.87}/\second{84.81}
& \second{86.85}/\second{86.91} & 70.47/69.94 & \second{85.80}/\second{85.68} & \second{87.62}/\second{87.65} \\
& MMIN
& 62.78/61.92 & 50.87/\second{46.11} & 78.59/78.57
& 82.19/82.24 & 70.07/69.82 & 80.77/80.65 & 83.67/83.64 \\
& LNLN
& 57.64/55.82 & 49.28/42.90 & 78.69/78.66
& 81.92/82.04 & 66.97/66.58 & 80.59/80.42 & 83.68/83.71 \\
& MCULoRA
& 61.19/59.44 & 46.75/38.43 & 80.09/80.12
& 81.91/82.10 & \second{72.47}/\second{72.37} & 82.54/82.28 & 85.44/85.46 \\
& ComP
& 52.02/48.71 & \second{51.41}/43.12 & 80.12/80.18
& 83.34/83.44 & 68.83/68.21 & 83.31/83.15 & 85.72/85.67 \\
& \ourscell CoRe-KD
& \ourscell \best{76.17}/\best{75.83}
& \ourscell \best{61.43}/\best{60.32}
& \ourscell \best{84.94}/\best{84.90}
& \ourscell \best{88.73}/\best{88.67}
& \ourscell \best{80.68}/\best{80.79}
& \ourscell \best{86.55}/\best{86.44}
& \ourscell \best{89.21}/\best{89.03} \\
& \ourscell $\Delta$SOTA
& \ourscell $\uparrow$ 7.56/7.63
& \ourscell $\uparrow$ 10.02/14.21
& \ourscell $\uparrow$ 0.07/0.09
& \ourscell $\uparrow$ 1.88/1.76
& \ourscell $\uparrow$ 8.21/8.42
& \ourscell $\uparrow$ 0.75/0.76
& \ourscell $\uparrow$ 1.59/1.38 \\
\hline

\multicolumn{2}{c||}{Missing Rate}
& 0.1 & 0.2 & 0.3 & 0.4 & 0.5 & 0.6 & 0.7 \\
\hline
Dataset & Method
& Acc(\%)/F1(\%) & Acc(\%)/F1(\%) & Acc(\%)/F1(\%)
& Acc(\%)/F1(\%) & Acc(\%)/F1(\%) & Acc(\%)/F1(\%)
& Acc(\%)/F1(\%) \\
\hline

\multirow{8}{*}{\begin{tabular}{c}IEMOCAP\\Four\end{tabular}}
& IMDer
& 81.79/81.80 & 80.37/80.41 & 79.37/79.43
& 77.68/77.75 & 75.25/75.33 & 71.95/72.00 & 66.31/66.29 \\
& Corr-KD
& \second{87.11}/\second{87.15} & \second{86.24}/\second{86.28} & \second{85.58}/\second{85.61}
& \second{84.07}/\second{84.09} & \second{83.04}/\second{83.06} & \second{81.80}/\second{81.77} & \second{80.06}/\second{79.89} \\
& MMIN
& 82.68/82.64 & 81.12/81.11 & 79.88/79.89
& 78.64/78.66 & 76.36/76.39 & 73.39/73.41 & 68.47/68.41 \\
& LNLN
& 82.70/82.70 & 81.15/81.18 & 80.09/80.13
& 78.63/78.69 & 76.43/76.46 & 73.48/73.41 & 68.67/68.46 \\
& MCULoRA
& 84.35/84.37 & 83.42/83.44 & 81.91/81.93
& 80.73/80.75 & 78.41/78.41 & 76.86/76.81 & 73.62/73.53 \\
& ComP
& 84.63/84.58 & 83.69/83.65 & 82.37/82.33
& 81.08/81.03 & 78.80/78.75 & 76.28/76.25 & 72.50/72.45 \\
& \ourscell CoRe-KD
& \ourscell \best{88.73}/\best{88.63}
& \ourscell \best{88.33}/\best{88.22}
& \ourscell \best{87.12}/\best{86.99}
& \ourscell \best{86.23}/\best{86.09}
& \ourscell \best{86.72}/\best{86.63}
& \ourscell \best{85.02}/\best{84.90}
& \ourscell \best{83.98}/\best{83.91} \\
& \ourscell $\Delta$SOTA
& \ourscell $\uparrow$ 1.62/1.48
& \ourscell $\uparrow$ 2.09/1.94
& \ourscell $\uparrow$ 1.54/1.38
& \ourscell $\uparrow$ 2.16/2.00
& \ourscell $\uparrow$ 3.68/3.57
& \ourscell $\uparrow$ 3.22/3.13
& \ourscell $\uparrow$ 3.92/4.02 \\
\Xhline{0.9pt}
\end{tabular}%
}

\caption{
Supplementary results on IEMOCAP-4 under fixed-missing and random-missing protocols.
Best results are bolded and second-best results are underlined.
}
\label{tab:iemocap4_missing_results}
\end{table*}

We report supplementary IEMOCAP-4 results under fixed-missing and random-missing protocols.
This setting follows the commonly used four-class IEMOCAP protocol and is included to provide additional comparison with prior conversational MER studies.
As shown in Table~\ref{tab:iemocap4_missing_results}, CoRe-KD achieves the best Acc/F1 in all fixed-missing conditions and all random-missing rates.
The gains are especially large when only acoustic or visual observations are available, and also increase under higher random-missing rates.
This pattern is consistent with the main IEMOCAP-6 results, suggesting that complete-view state anchoring provides robust supervision across different IEMOCAP label protocols.
This table provides the per-condition source results for the IEMOCAP-4 setting, complementing the averaged scores in Table~\ref{tab:seed_variance_all}.

\section{Full Ablation}
\label{app:full_ablation}

\begin{table*}[t]
\centering
\small
\setlength{\tabcolsep}{3.0pt}
\renewcommand{\arraystretch}{1.04}
\setlength{\arrayrulewidth}{0.4pt}
\setlength{\doublerulesep}{1.2pt}

\providecommand{\abph}{xx.xx/xx.xx}
\providecommand{\abours}{\cellcolor{gray!12}}
\providecommand{\best}[1]{\textbf{#1}}

\resizebox{0.99\textwidth}{!}{%
\begin{tabular}{@{}c|l||*{7}{c}@{}}
\Xhline{0.9pt}
\multicolumn{2}{c||}{Testing Condition}
& $\{\mathrm a\}$ & $\{\mathrm v\}$ & $\{\mathrm \ell\}$
& $\{\mathrm \ell,\mathrm a\}$ & $\{\mathrm a,\mathrm v\}$
& $\{\mathrm \ell,\mathrm v\}$ & $\{\mathrm \ell,\mathrm a,\mathrm v\}$ \\
\hline
Dataset & Variant
& Acc(\%)/F1(\%) & Acc(\%)/F1(\%) & Acc(\%)/F1(\%)
& Acc(\%)/F1(\%) & Acc(\%)/F1(\%) & Acc(\%)/F1(\%)
& Acc(\%)/F1(\%) \\
\hline

\multirow{6}{*}{\begin{tabular}{c}IEMOCAP\\Four\end{tabular}}
& w/o $\mathcal{L}_{\mathrm{CSA}}$
& 69.65/68.81 & 49.76/45.85 & 82.21/82.24 & 86.39/86.44 & 72.54/71.45 & 84.38/84.16 & 87.84/87.72 \\

& w/o $\mathcal{L}_{\mathrm{NCE}}$
& 68.28/67.91 & 45.57/44.24 & 83.49/83.14 & 85.75/85.60 & 72.54/72.23 & 84.30/83.93 & 85.19/84.93 \\

& w/o $\mathcal{L}_{\mathrm{pred}}$
& 72.06/72.07 & 51.05/48.49 & 83.90/83.79 & 87.04/87.10 & 78.10/77.95 & 84.70/84.49 & 87.52/87.48 \\

& w/o $\mathcal{L}_{\mathrm{state}}$
& 74.48/74.42 & 53.46/51.26 & 82.53/82.62 & 87.20/87.28 & 80.35/80.18 & 84.78/84.71 & 88.16/88.15 \\

& w/o $\mathcal{L}_{\mathrm{mstate}}$
& 75.52/75.18 & 50.00/46.93 & 83.74/83.78 & 87.12/87.15 & 78.18/77.61 & 84.70/84.61 & 88.08/88.02 \\

& \abours CoRe-KD
& \abours \best{76.17}/\best{75.83}
& \abours \best{61.43}/\best{60.32}
& \abours \best{84.94}/\best{84.90}
& \abours \best{88.73}/\best{88.67}
& \abours \best{80.68}/\best{80.79}
& \abours \best{86.55}/\best{86.44}
& \abours \best{89.21}/\best{89.03} \\
\hline

\multirow{6}{*}{\begin{tabular}{c}MELD\\Seven\end{tabular}}
& w/o $\mathcal{L}_{\mathrm{CSA}}$
& 18.62/24.07 & 26.74/29.82 & 61.65/63.86 & 60.34/63.41 & 19.66/24.87 & 60.84/63.46 & 59.08/62.30 \\

& w/o $\mathcal{L}_{\mathrm{NCE}}$
& 43.33/40.56 & 29.89/29.77 & 66.05/66.33 & 65.63/66.02 & 37.13/37.15 & 65.36/66.00 & 65.52/66.24 \\

& w/o $\mathcal{L}_{\mathrm{pred}}$
& 22.15/25.50 & 29.50/32.00 & 60.73/63.26 & 60.23/63.08 & 24.60/29.27 & 61.00/63.52 & 59.43/62.55 \\

& w/o $\mathcal{L}_{\mathrm{state}}$
& 48.12/31.27 & 48.12/31.27 & 63.60/58.70 & 63.37/58.18 & 48.12/31.27 & 62.80/57.00 & 63.03/57.15 \\

& w/o $\mathcal{L}_{\mathrm{mstate}}$
& 48.12/31.27 & 48.12/31.27 & 54.94/42.53 & 55.33/42.83 & 48.12/31.27 & 55.17/42.70 & 55.21/42.73 \\

& \abours CoRe-KD
& \abours \best{49.77}/\best{41.08}
& \abours \best{48.35}/\best{32.24}
& \abours \best{68.28}/\best{67.25}
& \abours \best{68.47}/\best{67.35}
& \abours \best{50.19}/\best{38.59}
& \abours \best{68.66}/\best{67.97}
& \abours \best{68.85}/\best{67.41} \\
\hline

\multirow{6}{*}{\begin{tabular}{c}CMU\\MOSEI\end{tabular}}
& w/o $\mathcal{L}_{\mathrm{CSA}}$
& 72.04/70.35 & 58.34/58.41 & 85.09/85.25 & 85.97/86.06 & 71.90/71.79 & 84.95/85.14 & 85.94/86.08 \\

& w/o $\mathcal{L}_{\mathrm{NCE}}$
& 70.72/69.59 & 66.68/64.33 & 85.53/85.67 & 85.58/85.74 & 70.06/69.82 & 86.10/86.23 & 85.61/85.76 \\

& w/o $\mathcal{L}_{\mathrm{pred}}$
& 71.93/69.70 & 51.16/48.99 & 85.42/85.57 & 85.77/85.90 & 71.52/71.49 & 84.78/84.99 & 85.64/85.80 \\

& w/o $\mathcal{L}_{\mathrm{state}}$
& 66.81/59.69 & 56.99/57.17 & 83.54/83.81 & 84.89/85.09 & 71.00/69.02 & 83.16/83.43 & 84.76/84.97 \\

& w/o $\mathcal{L}_{\mathrm{mstate}}$
& 70.36/66.63 & 53.41/52.44 & 86.21/86.29 & 86.90/86.90 & 72.73/72.32 & 85.94/86.09 & 86.60/86.69 \\

& \abours CoRe-KD
& \abours \best{72.95}/\best{72.39}
& \abours \best{67.01}/\best{67.24}
& \abours \best{86.49}/\best{86.49}
& \abours \best{86.98}/\best{86.98}
& \abours \best{72.81}/\best{72.67}
& \abours \best{86.82}/\best{86.85}
& \abours \best{87.12}/\best{87.16} \\
\hline

\multicolumn{2}{c||}{Missing Rate}
& 0.1 & 0.2 & 0.3 & 0.4 & 0.5 & 0.6 & 0.7 \\
\hline
Dataset & Variant
& Acc(\%)/F1(\%) & Acc(\%)/F1(\%) & Acc(\%)/F1(\%)
& Acc(\%)/F1(\%) & Acc(\%)/F1(\%) & Acc(\%)/F1(\%)
& Acc(\%)/F1(\%) \\
\hline

\multirow{6}{*}{\begin{tabular}{c}IEMOCAP\\Four\end{tabular}}
& w/o $\mathcal{L}_{\mathrm{CSA}}$
& 87.44/87.32 & 86.72/86.60 & 86.15/86.04 & 85.51/85.39 & 85.02/84.88 & 82.69/82.48 & 81.48/81.21 \\

& w/o $\mathcal{L}_{\mathrm{NCE}}$
& 85.67/85.49 & 85.10/84.95 & 83.98/83.80 & 83.41/83.27 & 83.17/83.01 & 82.05/81.85 & 80.60/80.34 \\

& w/o $\mathcal{L}_{\mathrm{pred}}$
& 87.04/86.99 & 86.80/86.74 & 85.91/85.82 & 85.19/85.09 & 84.38/84.28 & 83.82/83.67 & 82.13/81.87 \\

& w/o $\mathcal{L}_{\mathrm{state}}$
& 87.04/87.02 & 86.96/86.93 & 85.51/85.48 & 85.02/84.97 & 84.38/84.32 & 82.93/82.87 & 82.13/81.96 \\

& w/o $\mathcal{L}_{\mathrm{mstate}}$
& 87.52/87.44 & 87.20/87.12 & 86.63/86.56 & 85.67/85.60 & 85.75/85.64 & 84.22/84.08 & 82.85/82.66 \\

& \abours CoRe-KD
& \abours \best{88.73}/\best{88.63}
& \abours \best{88.33}/\best{88.22}
& \abours \best{87.12}/\best{86.99}
& \abours \best{86.23}/\best{86.09}
& \abours \best{86.72}/\best{86.63}
& \abours \best{85.02}/\best{84.90}
& \abours \best{83.98}/\best{83.91} \\
\hline

\multirow{6}{*}{\begin{tabular}{c}MELD\\Seven\end{tabular}}
& w/o $\mathcal{L}_{\mathrm{CSA}}$
& 56.13/59.80 & 53.33/57.55 & 49.31/54.26 & 45.52/50.93 & 42.30/48.34 & 37.82/44.35 & 31.65/38.29 \\

& w/o $\mathcal{L}_{\mathrm{NCE}}$
& 63.10/63.72 & 61.00/61.63 & 58.28/58.80 & 54.83/55.31 & 52.61/52.95 & 50.15/50.58 & 47.13/47.33 \\

& w/o $\mathcal{L}_{\mathrm{pred}}$
& 56.90/60.21 & 54.52/58.30 & 50.15/54.37 & 47.43/52.00 & 43.41/48.29 & 39.43/44.35 & 35.13/40.11 \\

& w/o $\mathcal{L}_{\mathrm{state}}$
& 61.65/55.32 & 60.19/53.43 & 58.35/51.18 & 56.48/48.47 & 54.83/45.95 & 53.60/43.86 & 51.92/40.62 \\

& w/o $\mathcal{L}_{\mathrm{mstate}}$
& 54.52/42.00 & 53.83/41.21 & 53.14/40.41 & 52.38/39.46 & 51.95/38.77 & 51.19/37.71 & 50.27/36.14 \\

& \abours CoRe-KD
& \abours \best{66.36}/\best{65.56}
& \abours \best{64.18}/\best{63.28}
& \abours \best{61.84}/\best{60.70}
& \abours \best{60.11}/\best{57.61}
& \abours \best{58.81}/\best{56.55}
& \abours \best{56.32}/\best{52.78}
& \abours \best{54.18}/\best{48.42} \\
\hline

\multirow{6}{*}{\begin{tabular}{c}CMU\\MOSEI\end{tabular}}
& w/o $\mathcal{L}_{\mathrm{CSA}}$
& 85.17/85.30 & 84.15/84.27 & 82.91/83.02 & 82.36/82.41 & 80.74/80.72 & 78.92/78.77 & 77.52/77.07 \\

& w/o $\mathcal{L}_{\mathrm{NCE}}$
& 84.70/84.82 & 83.68/83.76 & 82.80/82.82 & 82.36/82.24 & 81.07/80.76 & 80.13/79.59 & 77.88/76.72 \\

& w/o $\mathcal{L}_{\mathrm{pred}}$
& 84.98/85.14 & 83.76/83.94 & 82.09/82.27 & 81.70/81.84 & 79.99/80.12 & 79.47/79.54 & 77.85/77.74 \\

& w/o $\mathcal{L}_{\mathrm{state}}$
& 83.85/84.05 & 82.80/83.01 & 82.17/82.36 & 81.48/81.60 & 80.35/80.41 & 79.25/79.23 & 77.85/77.61 \\

& w/o $\mathcal{L}_{\mathrm{mstate}}$
& 85.80/85.87 & 84.51/84.54 & 83.19/83.19 & 83.02/82.90 & 81.04/80.79 & 79.53/78.98 & 77.82/76.71 \\

& \abours CoRe-KD
& \abours \best{86.49}/\best{86.40}
& \abours \best{85.06}/\best{84.89}
& \abours \best{84.23}/\best{84.04}
& \abours \best{83.52}/\best{83.21}
& \abours \best{82.00}/\best{81.54}
& \abours \best{81.10}/\best{80.42}
& \abours \best{78.92}/\best{77.77} \\
\Xhline{0.9pt}
\end{tabular}%
}
\caption{
Extended ablation table of CoRe-KD on IEMOCAP-4, MELD-7, and CMU-MOSEI under fixed-missing testing conditions and random-missing rates.
Ablation rows report the corresponding component-removal results, while the CoRe-KD rows are kept aligned with the main-table and supplementary-table results.
}
\label{tab:ablation_iemocap4_meld_mosei}
\end{table*}

Table~\ref{tab:ablation_iemocap4_meld_mosei} extends the component ablation results to IEMOCAP-4, MELD-7, and CMU-MOSEI.
The CoRe-KD rows are aligned with the corresponding main-table and supplementary-table results.
Across these additional settings, the full model remains strongest in most fixed- and random-missing conditions.
Removing \(\mathcal L_{\mathrm{CSA}}\) or \(\mathcal L_{\mathrm{pred}}\) often leads to large drops, especially in low-information conditions where only nonverbal modalities are available.
Removing \(\mathcal L_{\mathrm{state}}\), \(\mathcal L_{\mathrm{mstate}}\), or \(\mathcal L_{\mathrm{NCE}}\) also weakens performance in many cases, suggesting that fused-state alignment, unavailable-modality-state anchoring, and nonverbal conflict regularization provide complementary gains.
These extended results provide additional checks that the component-wise trends observed on IEMOCAP-6 are not restricted to a single dataset setting.
\section{Computational Cost}
\label{app:cost}

CoRe-KD introduces additional computation only during training.
The complete-view teacher contains 84.8M parameters and is trained once before student training.
During student training, teacher references and NCE conflict views increase the training time from 0.14 hours to 0.40 hours on a single NVIDIA RTX A5000 GPU, corresponding to an additional 183.3\% training cost.
At inference time, the complete-view teacher, unavailable-state decoders, auxiliary training heads, and NCE donor construction are discarded.
The deployed model contains 88.0M parameters, where the student-side state heads and PoE fusion module introduce only 8.3M additional parameters, accounting for 9.4\% of the deployed model.
The inference latency is 30.3 ms per batch, compared with 28.8 ms for the baseline student under the same batch size.
Therefore, CoRe-KD introduces no teacher-side computation at deployment time, and its remaining overhead is limited to the lightweight state heads and PoE fusion used by the student.

\section{Additional t-SNE Visualization under Random Missing}
\label{app:tsne_rmfm}

To provide a qualitative view of representation robustness, we further visualize the fused representations on the IEMOCAP-6 test set under random modality feature missing (RMFM) in one seed.
The RMFM rate is varied from 0.1 to 0.7, where larger rates indicate more severe sample-wise modality incompleteness.
For each RMFM rate, we compare Corr-KD, LNLN, MoMKE, and CoRe-KD under the same visualization protocol.
Each point denotes one test utterance and is colored by its ground-truth emotion label.
The weighted F1 score of each setting is also reported in the corresponding panel.

\begin{figure*}[p]
    \centering
    \includegraphics[
        width=0.98\textwidth,
        height=0.88\textheight,
        keepaspectratio
    ]{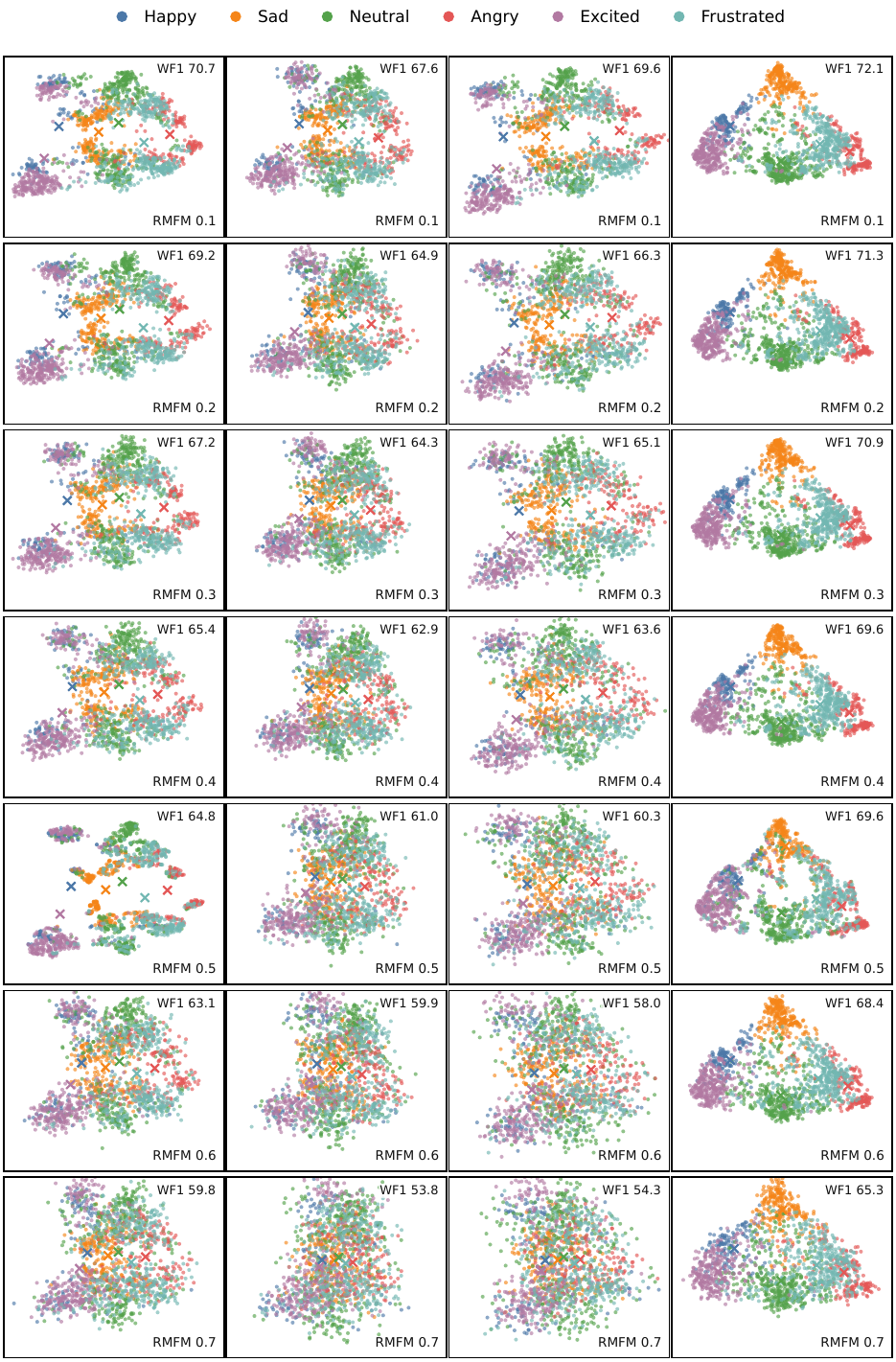}
    \caption{
    \textbf{t-SNE visualization on IEMOCAP-6 under random modality feature missing.}
    Rows correspond to RMFM rates from 0.1 to 0.7, and columns correspond to Corr-KD, LNLN, MoMKE, and CoRe-KD, respectively.
    Each point denotes one test utterance and is colored by its ground-truth emotion label.
    Each panel reports the corresponding weighted F1 score.
    The visualization provides a qualitative supplement to the quantitative RMFM results by showing how the fused representation space changes as modality incompleteness becomes more severe.
    }
    \label{fig:tsne_rmfm_iemocap6}
\end{figure*}

As shown in Figure~\ref{fig:tsne_rmfm_iemocap6}, the baseline methods tend to exhibit increasingly mixed class layouts as the missing rate increases, especially under high RMFM rates such as 0.5, 0.6, and 0.7.
In contrast, CoRe-KD maintains a more coherent representation structure across different missing rates, with visually clearer class regions and less severe mixing among several emotion categories.
This trend is consistent with the RMFM performance reported in the main results, where CoRe-KD degrades more slowly under increasing modality incompleteness.
These qualitative results suggest that complete-view state anchoring helps the student preserve a more stable fused state space under incomplete observations, while the conflict-aware training objective further regularizes the model against unreliable nonverbal evidence.
We emphasize that this visualization is used as a qualitative diagnostic rather than independent proof of representation separability.

\section{Ethics and Limitations}
\label{app:ethics}

CoRe-KD is evaluated on existing public emotion-recognition datasets and does not introduce new data collection.
However, emotion recognition remains a sensitive application area.
Emotion labels are subjective and may depend on culture, context, annotator background, and speaker identity.
Therefore, strong performance on benchmark splits should not be interpreted as reliable recognition of internal emotional states in real-world deployment.

Robustness under missing modalities can make deployment in low-quality sensing conditions more tempting.
This may increase the risk of overclaiming reliability when language, acoustic, or visual signals are noisy, incomplete, or socially ambiguous.
In addition, nonverbal expressions vary across speakers, communities, cultures, and disabilities.
A model trained to suppress conflicting nonverbal evidence may inadvertently treat some communication styles as unreliable or noisy.

For these reasons, CoRe-KD should not be used for high-stakes profiling, surveillance, hiring, workplace monitoring, education monitoring, or other consequential decision-making applications without strong safeguards, domain validation, and human oversight.
The method is intended as a research contribution for robust multimodal learning under controlled benchmark settings.

\end{document}